\documentclass[a4paper,superscriptaddress,floatfix]{revtex4}

\usepackage{graphicx}
\usepackage{subcaption}
\usepackage{endnotes}

\usepackage[utf8]{inputenc}
\usepackage[T1]{fontenc}
\usepackage[english]{babel}
\usepackage{blindtext}

\usepackage{caption}
\usepackage{slashed}
\usepackage{ragged2e}
\usepackage{pstricks}
\usepackage{color}
\usepackage{hepnames}
\usepackage{feynmp}
\usepackage{tikz}

\usepackage{amsmath}
\usepackage{amsfonts}
\usepackage{amssymb}
\usepackage{fancyhdr}
\usepackage{setspace}
\usepackage{bm}
\usepackage{units}
\usepackage{lineno}
\usepackage{multirow}
\usepackage{textcomp}

\usepackage{ifpdf}
\ifpdf
\fi

\DeclareUnicodeCharacter{03B3}{$\gamma$} 

\makeatletter
\def\endfmffile{%
	\fmfcmd{\p@rcent\space the end.^^J%
		end.^^J%
		endinput;}%
	\if@fmfio
	\immediate\closeout\@outfmf
	\fi
	\IfFileExists{\thefmffile.mp}{\immediate\write18{mpost \thefmffile}}{}
	\let\thefmffile\relax
}
\makeatother

\usepackage{hyperref}
\usepackage{doi}
\usepackage{url}

\usepackage{flafter}
\usepackage{placeins}
\usepackage{float}

\usepackage{lmodern}

\usepackage{tabularx}

\usepackage{pifont}

\usepackage{enumitem}

\newcommand{\explain}[2]{\underbrace{#1}_{\scriptscriptstyle\raggedright \textrm{#2}}}

\newcommand{\cba}          {\ensuremath{\cos(\beta - \alpha)}\xspace}
\newcommand{\tb}           {\ensuremath{\tan\beta}\xspace}

\newcommand{\Magellan}{\texttt{Magellan}\xspace}
\newcommand{\FeynRules}{\texttt{FeynRules}\xspace}

\newcommand{\MadGraph}{\texttt{MadGraph5}\xspace}
\newcommand{\MadAnalysis}{\texttt{MadAnalysis}\xspace}
\newcommand{\TTPS}{\texttt{T3PS}\xspace}
\newcommand{\HiggsSignals}{\texttt{HiggsSignals}\xspace}
\newcommand{\HiggsBounds}{\texttt{HiggsBounds}\xspace}
\newcommand{\THDMC}{\texttt{2HDMC}\xspace}
\newcommand{\Gfitter}{\texttt{Gfitter}\xspace}

\newcommand{\GeV}{\ensuremath{\mathrm{GeV}}\xspace}
\newcommand{\Delph}{\texttt{Delphes}\xspace}
\newcommand{\fast}{\texttt{FastJet}\xspace}

\begin{document}

\title{Exploring SM-like Higgs Boson Production in Association with Single-Top \\ 
at the LHC Within a 2HDM}

\author{C. Byers}
\email[E-mail: ]{c.byers@soton.ac.uk}
\affiliation{School of Physics \& Astronomy, University of Southampton, Highfield, Southampton SO17 1BJ, UK}

\author{S. Jain}
\email[E-mail: ]{s.jain@soton.ac.uk}
\affiliation{School of Physics \& Astronomy, University of Southampton, Highfield, Southampton SO17 1BJ, UK}

\author{S. Moretti}
\email[E-mail: ]{s.moretti@soton.ac.uk; stefano.moretti@stfc.ac.uk; stefano.moretti@physics.uu.se}
\affiliation{School of Physics \& Astronomy, University of Southampton, Highfield, Southampton SO17 1BJ, UK}
\affiliation{Particle Physics Department, Rutherford Appleton Laboratory, Chilton, Didcot, Oxon OX11 0QX, UK}
\affiliation{Department of Physics \& Astronomy, Uppsala University, Box 516, SE-751 20 Uppsala, Sweden}

\author{E. Olaiya}
\email[E-mail: ]{emmanuel.olaiya@stfc.ac.uk}
\affiliation{Particle Physics Department, Rutherford Appleton Laboratory, Chilton, Didcot, Oxon OX11 0QX, UK}

\begin{abstract}
		We investigate the possibility of detectable 2-Higgs Doublet Model (2HDM) type-II cross-sections at the High-Luminosity phase of the Large Hadron Collider (HL-LHC) for the production of the Standard Model (SM)-like Higgs boson ($h$) in association with a single top (anti)quark over the parameter space region  corresponding to the so-called   `wrong-sign solution' of the bottom (anti)quark Yukawa coupling. We isolate the latter by using  the toolbox \Magellan, which performs Markov Chain Monte Carlo (MCMC) scans  in the presence of all current  theoretical and experimental constraints, which   relevance is accounted for accurately by using built-in Bayesian statistical methods.   It is found that the allowed points in the 2HDM type-II parameter space of the aforementioned kind would not only provide inclusive rates considerably above those of the equivalent SM process but also distributions in several kinematical observables that are very different from the SM, both of which can help disentangle the SM from the 2HDM hypothesis. This difference is a consequence of the bottom-gluon fusion sub-process, which in the 2HDM becomes dominant over all others,   
with the latter  remaining very close to the SM yields. We prove that this phenomenology would be observable at the HL-LHC for the illustrative example of $h\to b\bar b$ decays.  
\end{abstract}

\maketitle

\newpage
\section{Introduction}

Following the discovery of a Higgs boson in 2012 at the LHC, $h$, the focus of much of the particle physics community has been
on measuring its properties: mass, width, quantum numbers, couplings, etc. Altogether, the discovered Higgs state is consistent with the one predicted by the SM (hence, it is very SM-like in its nature). However, when it comes to measuring couplings (which at present include those to $W^\pm, Z$ bosons and $t,b,c,\tau,\mu$ fermions), it is fair to say that much of the sensitivity is to the modulus of these, as the production and decay processes employed to perform these measurements are such that interference between these couplings amongst themselves or with others within the SM are non-existent or generally negligible. In fact, some degrees of access to the signs of couplings only really occur in $h\to \gamma\gamma$ and $h\to Z\gamma$ decays, wherein top (anti)quarks and charged weak bosons enter simultaneously at (the same) loop level. However, what is notably missing is access to the sign of the Higgs boson to bottom (anti)quark coupling. This is because the Yukawa nature of the SM-like Higgs couplings to fermions implies that the $ht\bar t$ strength is much higher than the $hb\bar b$ one, so that in the aforementioned decay processes the role of the latter is negligible in comparison to that of the former (This is also true in the production process $gg\to h$)\footnote{}\footnotetext[1]{A rather up-to-date review of the current LHC status regarding establishing the nature of the SM-like Higgs boson can be found in Ref.~\cite{Khalil:2022toi}.}.

\begin{figure}[H]
\begin{center}
    \includegraphics[width=0.8\linewidth]{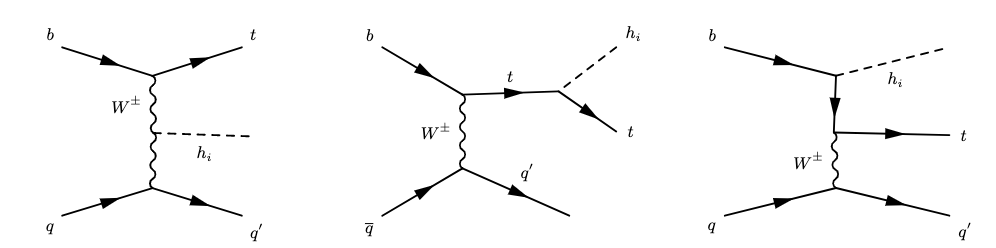}
\end{center}
		\caption{\label{Feyn_1} Feynman diagrams for the bq sub-process, assuming time flowing rightwards, wherein we ignore the contribution of a charged Higgs boson ($H^\pm$), which we take heavy enough so as to give a negligible correction. Notice that same diagrams appear in the qq sub-process, when time is flowing upwards.} 

\end{figure}

\begin{figure}[H]
\begin{center}
     \includegraphics[width=0.8\linewidth]{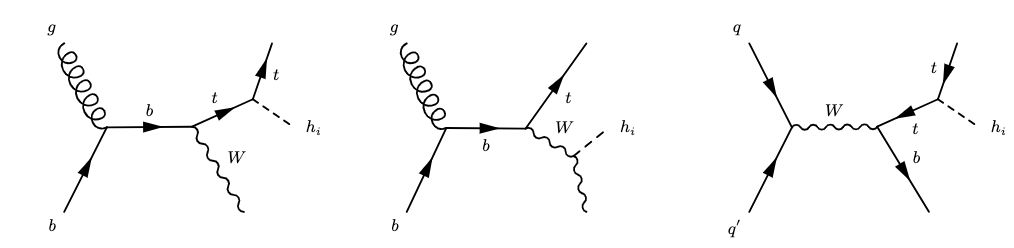}
\end{center}
		\caption{\label{Feyn_2} Feynman diagrams for the bg sub-process, assuming time flowing rightwards, wherein we ignore the contribution of a charged Higgs boson ($H^\pm$), which we take heavy enough so as to give a negligible correction.} 

\end{figure}

This is altogether an advantage, though, if one assumes  some Beyond the SM (BSM) construct as the one governing the Higgs dynamics, since, in some realisations of the latter, the  $hb\bar b$ coupling can have an opposite sign with respect to the SM. On the one hand, current measurements would not be affected by this. On the other hand, alternative measurements might well be. In this context, it becomes important to exploit other $h$ boson production channels in addition to the customary ones that have already been established at the LHC\footnote{}\footnotetext[2]{Hereafter, '$q^{(')}$' refers to a light quark ($d,u,s$ or $c$). }: gluon-gluon fusion
($gg\to h$), vector-boson fusion ($qq\to q^{(')}q^{(')}h$) and associated production with weak gauge bosons ($q\bar q^{(')}\to Zh(W^\pm h)$) or top (anti)quark pairs ($q\bar q,gg\to t\bar t h$) (see Ref.~\cite{Kunszt:1996yp} for a review). Specifically, we concentrate here on SM-like Higgs boson production in association with a single top (anti)quark. The latter is mediated by the following sub-processes: $bq\to t q' h$ + c.c. (hereafter, bq), $bg \rightarrow t W^- h$ + c.c. (hereafter, bg) and $q\bar q'\to t\bar b h$ + c.c. (hereafter, qq), see Figs.~\ref{Feyn_1}--\ref{Feyn_2}. In the SM, the corresponding cross-sections at the LHC are in decreasing order of importance in relation to the order in which we have listed them, see Tab.~\ref{SM-Xsects}. Altogether, their production cross-section is smaller than but of the same order as that of $q\bar q,gg\to t\bar t h$, so some sensitivity presently exists to this additional $h$ production mechanism \cite{CMS-PAS-HIG-19-008,Sirunyan:2648886}. Furthermore, in  all such analyses, the bq,  bg and qq channels are treated inclusively.

Unfortunately, current searches have only been able to exclude cross-sections for SM-like Higgs boson production in association with a single top (anti)quark for values above the SM predictions. This is primarily owed to the fact that there exist cancellations between the topologies in Figs.~\ref{Feyn_1}--\ref{Feyn_2}, in turn, driven by the fact that, unlike the other four production processes mentioned above, herein, the $h$ couplings to $W^\pm$ bosons and $t,b$ quarks enter simultaneously at amplitude level and can therefore interfere.  As a consequence of all this, though, such a mechanism may be a privileged probe of some BSM dynamics, as any alteration of these three Higgs couplings may lead to larger cross-sections than SM predictions, thereby rendering the mechanism observable at the HL-LHC, sooner than otherwise expected when only considering the SM. Furthermore, as the Feynman diagrams carrying such couplings are topologically different (i.e., they would produce a different kinematics in the final state), it may well be that differential distributions in some BSM scenario may also be different from those produced in the SM case.

With this in mind, in this paper, we study the possibility of all such a phenomenology being realised in the simplest extension of the SM Higgs sector using doublet fields (like in the SM), i.e., the one embedded in a generic 2HDM \cite{Branco2012}. In particular, we consider the $h\to b\bar b$ decay channel, as it strikes a good balance of expecting of a substantial number of events to be reconstructed and a manageable background when in conjunction with the other signal final state particles. This is one of the dominant decay channels amongst those of the SM, yet it is poorly measured because of the large background arising whenever this is searched for through the standard four production processes\footnote{}\footnotetext[3]{Indeed, current estimates point to the extraction of a signal for SM-like Higgs boson production in association with a single top (anti)quark  in the SM being possible  with a minimum of 1500 fb$^{-1}$ of integrated  luminosity in the $h\to b\bar{b}$ channel\cite{Kobakhidze:2014gqa,Farina:2012xp}, which is only achievable at the High-Luminosity LHC (HL-LHC).}. We will be showing that a significant increase in cross-section in the 2HDM type-II  is possible, in particular, for the bg sub-process, while the bq and qq rates  remain close to their SM counterparts. Remarkably, this happens precisely when the $hb\bar b$ coupling changes sign with respect to the SM, which happens for the so-called `wrong-sign solution'  to the current SM-like Higgs boson measurements, which we have defined using the package \Magellan\ \cite{Accomando:2022nfc}. In fact, as intimated, given that the kinematics emerging from the diagram carrying the $hb\bar b$ coupling is different from that involving the $ht\bar t$ and $hW^+W^-$, we will also be able to show that significant differences exist in a variety of  differential distributions between the SM and 2HDM type-II cases. Altogether, this sets the stage for vigorously pursuing experimentally this additional $h$ production channel, with  a dual purpose in sight. On the one hand, to prove the existence of a BSM Higgs sector. On the other hand, to show that it should be possible to disentangle the underlying structure of it (at least in the 2HDM type-II).

\begin{table}[h!]
	\begin{center}
		\newcolumntype{C}[1]{>{\centering\let\newline\\\arraybackslash\hspace{0pt}}m{#1}} 
		\begin{tabular}{ |C{2cm}|C{2cm}| C{2.2cm} | C{2cm} | C{2cm}|}
			\hline
			& $\sigma({\rm bq})$ (pb)	& $\sigma({\rm bg})$ (pb)	& $\sigma({\rm qq})$ (pb) & $\sigma({\rm total})$ (pb)\\\hline
			SM	& 0.036 & 0.011 & 0.0023	& 0.049\textsuperscript{\cite{CMS-PAS-HIG-19-008}} \\
			\hline
		\end{tabular}
		\caption{\label{SM-Xsects} The tree-level cross-sections for the bq, bg and qq sub-processes of  SM-like Higgs boson production  in association with a single top (anti)quark at the LHC with 13.6 TeV of Centre-of-Mass (CM) energy. (These values have been calculated by \MadGraph-3.1.0 \cite{ostrolenk2021} for the default SM implementation that comes with the package.)}
	\end{center}
\end{table}
The plan of the paper is as follows. In the next section, we introduce  the 2HDM. This will be followed by a description of \Magellan and how we have performed parameter space scans using it. Then our detector-level analysis will be illustrated. We finally conclude in the last section. 

\section{The 2HDM}

Here we briefly describe the 2HDM, including its types, with a focus on the aspects relevant to our work. Extensive reviews of the 2HDM can be found in Refs.~\cite{Branco2012,HiggsHunters}. We consider the presence of a second Higgs doublet, giving us a generic 2HDM. A second Higgs doublet arises within many BSM physics models, including those originating  from  Supersymmetric theories. The additional sources of CP-violation present in this type of enlarged Higgs sector could provide an explanation for the apparent matter-antimatter asymmetry. Particular realisations of the 2HDM also have the appealing ability to explain neutrino mass generation~\cite{Neutrino-mass}, to provide dark matter candidates \cite{Ko2014} or to accommodate the muon $g-2$ anomaly \cite{muong2lepton, muong2limiting, Wang2015}. Additionally, it is well known that the Minimal Supersymmetric Standard Model (MSSM) requires the existence of two doublets to generate the mass of up-type quarks, down-type quarks and charged leptons.

An important feature of the 2HDM is the number of degrees of freedom (d.o.f) that the constituent fields have. These d.o.f. can be enumerated before and after the spontaneous breaking of the Electro-Weak (EW) symmetry due to the shape of the Higgs potential. Initially, we have a pair of complex doublets, $\Phi_{1}$ and $\Phi_{2}$, giving 8 d.o.f. in total. After EW Symmetry Breaking (EWSB), the spectrum contains two CP-even scalars $h$ and $H$, one pseudoscalar $A$ and two charged Higgs bosons $H^{\pm}$ (i.e. 5 d.o.f.). The Goldstone bosons of the theory will then become the longitudinal components of the weak $W^{\pm}$ and $Z$ bosons (3 d.o.f). Hence, the total d.o.f. number is unchanged.

The most general renormalisable, (i.e.,  quartic), the scalar potential of two doublets can be written as
\begin{equation}
\begin{aligned}
\mathcal{V}_{\rm gen}
&=
m_{11}^{2} \Phi_{1}^{\dag} \Phi_{1} 
+ m_{22}^{2} \Phi_{2}^{\dag} \Phi_{2}
- \left[ m_{12}^{2} \Phi_{1}^{\dag} \Phi_{2} + \textrm{h.c} \right] + \\
&+ \frac{1}{2} \lambda_{1} \left( \Phi_{1}^{\dag} \Phi_{1} \right)^{2}
+ \frac{1}{2} \lambda_{2} \left( \Phi_{2}^{\dag} \Phi_{2} \right)^{2}
+ \lambda_{3} \left( \Phi_{1}^{\dag} \Phi_{1} \right) \left( \Phi_{2}^{\dag} \Phi_{2} \right)
+ \lambda_{4} \left( \Phi_{1}^{\dag} \Phi_{2} \right) \left( \Phi_{2}^{\dag}
\Phi_{1} \right) + \\
&+ \left\{ \frac{1}{2} \lambda_{5} \left( \Phi_{1}^{\dag} \Phi_{2} \right)^{2} + \left[ \lambda_{6} \left( \Phi_{1}^{\dag} \Phi_{1} \right) + \lambda_{7} \left( \Phi_{2}^{\dag} \Phi_{2} \right) \right] \Phi_{1}^{\dag} \Phi_{2} \right\},
\end{aligned}
\end{equation}
\noindent
where $m_{11}^{2}$, $m_{22}^{2}$, $m_{12}^{2}$ are the mass squared parameters and $\lambda_{i}$ ($i = 1, ..., 7$) are dimensionless quantities describing the coupling of the order-4 interactions. Six of these parameters are real ($m_{11}^{2}$, $m_{22}^{2}$, $\lambda_{i}$ with $i = 1, ..., 4$) and four are a priori complex ($m_{12}^{2}$ and $\lambda_{i}$ with $i = 5, ..., 7$). Therefore, generally, the model has 14 free parameters. Under appropriate manipulations, however, it is possible to reduce this number.

The potential is explicitly CP-conserving if and only if there exists a basis choice for the scalar fields in which $m_{12}^{2}$, $\lambda_{5}$, $\lambda_{6}$ and $\lambda_{7}$ are real. Notice that the vacuum could still break CP spontaneously, even in this case. The spontaneous CP-violation of the vacuum takes place if and only if the scalar potential is explicitly CP-conserving, but there is no basis in which the scalars are real \cite{Haber2015}. In the following, we assume that both the scalar potential and vacuum are CP-conserving. By requiring CP-conservation, one looses 4 d.o.f. taking the number of free parameters down to 10. After EWSB, each scalar doublet acquires a Vacuum Expectation Value (VEV) that can be parametrised as follows:
\begin{equation}
	\langle \Phi_{1} \rangle = \frac{v}{\sqrt{2}}
	\begin{pmatrix}  
		0 \\
		\cos\beta
	\end{pmatrix}
	\quad
	\quad
	\langle \Phi_{2} \rangle = \frac{v}{\sqrt{2}}
	\begin{pmatrix}  
	0 \\
	\sin\beta
	\end{pmatrix},
\end{equation}
\noindent
where the angle $\beta$ is determined by the ratio of the two doublet VEVs, $v_1$ and $v_2$, through the definition of $\tan\beta = v_{2} / v_{1}$. Thus, $\beta$ is an additional parameter that must be added to the free parameters defining the scalar potential. 
{
	\begin{table}[t!]
		\centering
		\begin{tabular}{|c|c|c|c|c|c|c|c|c|c|}
			\hline
			\multirow{2}{*}{Model} & \multicolumn{3}{c|}{$h$} & \multicolumn{3}{c|}{$H$} & \multicolumn{3}{c|}{$A$} \tabularnewline
			\cline{2-10}
			&
			$u$ & $d$ & $l$ &
			$u$ & $d$ & $l$ &
			$u$ & $d$ & $l$
			\tabularnewline
			\hline
			{2HDM type-I}  &
			$\phantom{-}\frac{\cos\alpha}{\sin\beta}$ & $\phantom{-}\frac{\cos\alpha}{\sin\beta}$ & $\phantom{-}\frac{\cos\alpha}{\sin\beta}$ & 
			$\phantom{-}\frac{\sin\alpha}{\sin\beta}$ & $\phantom{-}\frac{\sin\alpha}{\sin\beta}$ & $\phantom{-}\frac{\sin\alpha}{\sin\beta}$ & 
			$\phantom{-}\cot\beta                   $ & $         - \cot\beta                   $ & $         - \cot\beta                $         
			\tabularnewline
			\hline
			{2HDM type-II}  &
			$\phantom{-}\frac{\cos\alpha}{\sin\beta}$ & $          -\frac{\sin\alpha}{\cos\beta} $ & $          -\frac{\sin\alpha}{\sin\beta} $ & 
			$\phantom{-}\frac{\sin\alpha}{\sin\beta}$ & $\phantom{-}\frac{\cos\alpha}{\sin\beta} $ & $\phantom{-}\frac{\cos\alpha}{\sin\beta} $ & 
			$            \cot\beta                  $ & $           \tan\beta                    $ & $           \tan\beta                    $   
			\tabularnewline
			\hline
		\end{tabular}
		\caption{
			Couplings of the neutral Higgs bosons to fermions, normalised to the corresponding SM
			value ($m_{f}/v$) in the 2HDM type-I and II.}
		\label{tab:interaction}
	\end{table}
}

In general, the Yukawa matrices corresponding to the two doublets are {not} simultaneously diagonalisable: this is a potential problem, as the off-diagonal elements lead to tree-level Higgs mediated Flavour Changing Neutral Currents (FCNCs) which have been sharply bounded by experiment. The Glashow-Weinberg-Paschos (GWP) theorem~\cite{GWP1,GWP2} states that such FCNCs are absent if at most one Higgs multiplet is responsible for providing mass to fermions of a given electric charge. This GWP condition can be enforced by a discrete $\mathbb{Z}_2$-symmetry (e.g., where $\Phi_1 \rightarrow +\Phi_1$ and $\Phi_2\rightarrow -\Phi_2$) on the doublets, in which case the lack of FCNCs come about in a natural way. The soft $\mathbb{Z}_2$ breaking condition relies on the existence of a basis wherin $\lambda_6 = \lambda_7$ = 0, with $m_{12}$ non-zero but small enough. Requiring soft breaking one then additionally  looses 2 d.o.f., taking the total number down to 9. Finally, $m_{11}^{2}$ and $m_{22}^{2}$ can both be expressed as a function of the other parameters: this is due to the fact that the scalar potential is in a local minimum when computed in the VEVs. So, altogether, with restrictions to CP-conservation and soft $\mathbb{Z}_2$-symmetry breaking applied, there remain 7 free parameters in the 2HDM.

The applied conditions allow for multiple bases in which the 2HDM can be described as follows: the \textit{general parametrisation} (as given above in terms of $m^2_{ij}$ and $\lambda_i$s), the \textit{Higgs basis}, where one of the doublets gets zero VEV, and the \textit{physical basis}, where one uses the physical masses of the (pseudo)scalars. However, in  the light of the discovery of the $125$ GeV Higgs boson, herein referred to as the $h$ state, it is common to parametrise the theory using the \textit{hybrid basis}~\cite{Haber2015}, where the parameters allow for direct control on both the CP-even and CP-odd Higgs masses, the $hVV$ couplings ($V$ =  $ W^\pm , Z$), the $Aq\bar{q}$ vertices and the Higgs quartic couplings. The parameters in this basis are:
\begin{equation}
\explain{
	m_{h}, \quad
	m_{H}}
{masses for the two CP-even Higgses}, \quad
\explain{\cos(\beta - \alpha)}{\text{\shortstack{ determines the \\ $g_{hVV}$ \& $g_{HVV}$ couplings }}}
, \quad
\explain{
	\tan\beta}
{given by the ratio of the vevs}
, \quad
\explain{
	Z_{4}, \quad
	Z_{5}, \quad
	Z_{7},}
{\text{\shortstack{self-coupling parameters\\ for the Higgses}}}
\end{equation}
\noindent
with $m_H\ge m_h$, $0\leq\beta\leq\pi /2$ and $0\le|\sin (\beta -\alpha )|\leq1$. The remaining (pseudo)scalar masses can be expressed in terms of the quartic scalar couplings in the Higgs basis:
\begin{equation}
m_A^2 = m_H^2\sin^2(\beta -\alpha) + m_h^2\cos^2(\beta -\alpha) - Z_5v_1^2,
\end{equation}
\begin{equation}
m_{H^\pm}^2 = m_A^2 -{\frac{1}{2}}(Z_4 - Z_5)v^2.
\end{equation}
\noindent
By swapping the self-couplings $Z_4$ and $Z_5$ with the scalar masses given above, it is possible to recast the 7 free parameters into 4 physical masses and 3 parameters related to the couplings of the (pseudo)scalars to gauge bosons, fermions and scalars themselves, respectively:
\begin{equation}
m_h, ~ m_H, ~ m_A, ~ m_{H^\pm}, ~ \cos (\beta - \alpha ), ~ \tan (\beta ), ~ Z_7.
\end{equation}
\noindent
It is worth noting that $Z_7$ only comes into play for the triple and quartic Higgs interactions, so it will not appear in the tree-level diagrams for our process. Finally, as $m_h$ has been measured to quite a degree of accuracy at the LHC, the number of d.o.f comes down to 6 globally.

Besides the (pseudo)scalar fields, fermions must also have a definite charge under the discrete $\mathbb{Z}_2$-symmetry. The different assignments of the $\mathbb{Z}_2$-charge in the fermion sector give rise to four different types of 2HDM, see \cite{Branco2012}. 
This paper will focus on 2HDM type-I and -II only. The couplings of the neutral Higgses to fermions, normalised to the corresponding SM value
($m_{f}/v$, henceforth, denoted by $\kappa_{hqq}$ or simply $\kappa_{qq}$ for the case of the SM-like Higgs state coupling to a quark $q$, where $q=d,u$), for these two types of 2HDM can be found in Tab.~\ref{tab:interaction}. 
When considering the type-II, there are two limiting scenarios, giving rise to two distinct regions in the 
$(\cos (\beta - \alpha), \tan \beta)$ parameter plane \cite{Ferreira2014}. These can be better understood by examining the behaviour of $\kappa_{hqq}$ as a function of the angles $\alpha$ and $\beta$. Taking the limits $\beta - \alpha\rightarrow \frac{\pi}{2}$ and $\beta + \alpha \rightarrow \frac{\pi}{2}$, the couplings become (recall  Tab.~\ref{tab:interaction}):
{
	\begin{equation}
	\begin{aligned}
	\kappa_{hdd}
	=
	- \frac{\sin \alpha}{ \cos \beta}
	&=
	\sin (\beta  - \alpha) - \cos( \beta - \alpha) \tan \beta 
	\xrightarrow[\beta - \alpha = \frac{\pi}{2}]{}
	1 ~ \textrm{(middle-region),}
	\\ 
	&=
	- \sin (\beta + \alpha) + \cos(\beta + \alpha ) \tan \beta
	\xrightarrow[\beta + \alpha = \frac{\pi}{2}]{}
	- 1 ~ \textrm{(right-arm),}
	\\
	\kappa_{huu}
	=
	\frac{\cos \alpha}{ \sin \beta}
	&=
	\sin (\beta  - \alpha) + \cos( \beta - \alpha) \cot \beta
	\xrightarrow[\beta - \alpha = \frac{\pi}{2}]{}
	1 ~ \textrm{(middle-region),}
	\\
	&=
	\sin (\beta + \alpha) + \cos(\beta + \alpha ) \cot \beta 
	\xrightarrow[\beta + \alpha = \frac{\pi}{2}]{}
	1 ~ \textrm{(right-arm).}
	\end{aligned}
	\end{equation}
}

\noindent

\begin{figure}[h!]
	\begin{center}
		\includegraphics[width=0.99\linewidth]{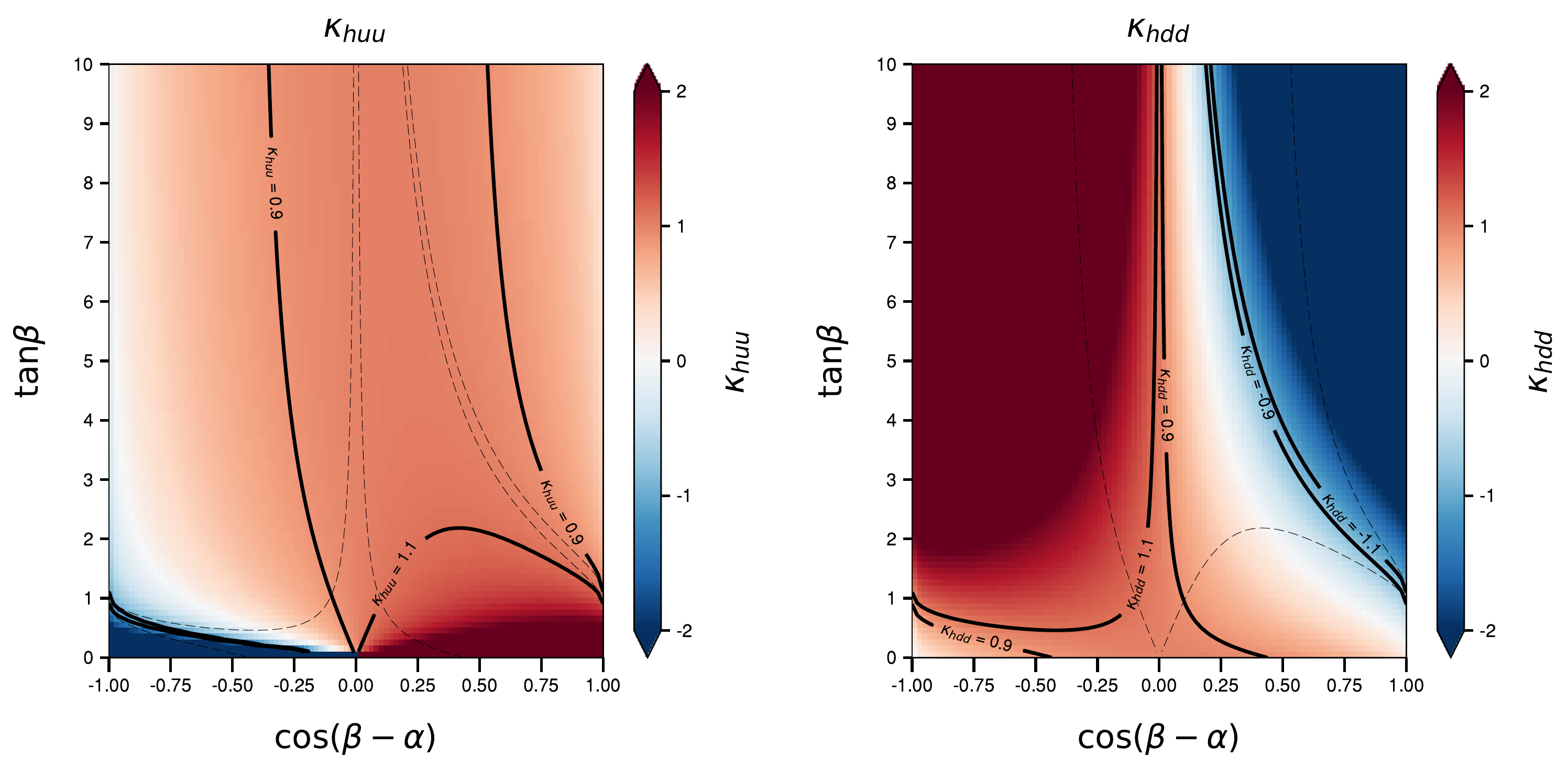}
		\caption{\label{fig:couplings}
			Light CP-even Higgs couplings to the up-type (left) and down-type (right) quarks, normalised to the corresponding
			SM value, in the ($\cos(\beta - \alpha), \tan \beta$) plane. Plots are taken directly from \cite{Accomando:2022nfc}.}
	\end{center}
\end{figure}

The dependence of $\kappa_{hdd}$ and $\kappa_{huu}$ on $\cos (\beta - \alpha )$ and 
$\tan (\beta )$ is illustrated in Fig.~\ref{fig:couplings}. The $\beta - \alpha \rightarrow \frac{\pi}{2}$ case corresponds to the ``middle-region'' and the SM-limit of the theory. The right-hand side plot of Fig.~\ref{fig:couplings}, is identified by the contour region where $0.9 \le\kappa_{hdd}\le 1.1$, this assumes a 10\% discrepancy from the SM values. The $\beta + \alpha\rightarrow \frac{\pi}{2}$ case corresponds to the ''right-arm'', where one gets an opposite sign for the coupling of the SM-like Higgs $h$ and the down-type quarks, relative to the SM value. This is known as the wrong-sign (Yukawa) coupling scenario. In the right-hand side plot of Fig.~\ref{fig:couplings}, we have this region represented by the narrow arm where the
coupling is negative. Again, this has a 10\% displacement from the SM value: $-1.1 \le \kappa_{hdd}\le -0.9$. Both the alignment and the wrong-sign regions are well within the O(10\%) discrepancy from their corresponding SM values which are allowed for the coupling of the SM-like Higgs to the up-type quarks, $\kappa_{huu}$, as shown in the left-hand plot of Fig.~\ref{fig:couplings}.

From an experimental point of view, the additional four states of a generic 2HDM~\cite{Branco2012,HiggsHunters} provide a range of observables through which a number of theoretical models could in principle be tested. Hence, it is worthwhile investigating in detail the scope of the LHC to discover new Higgs bosons as described within 2HDMs. Thus far there has been no experimental evidence for a 2HDM, but a vast array of literature exists on the phenomenological analyses which set bounds on the parameter space of such models. In the past two decades there has been considerable development and implementation of global fits, which collect the data coming from different experiments and perform rigorous statistical analyses to extract limits on BSM theories. The package \Gfitter \cite{Baak_2012} was a pioneer in releasing a global EW fit to constrain the New Physics (NP) predicted by a variety of models, including the 2HDM. Other such toolkits have been published in the literature, with their main focus centred on Supersymmetry and its variants (the 2HDM type-II in the decoupling limit being one such variant).

The standard procedure generally adopted by global fitting packages incorporates relevant experimental data and theoretical arguments that can constrain the parameter space of a NP model. These constraints can be categorised into the following three main sources: measurements of the discovered 125 \GeV Higgs boson properties (i.e., production and decay signal strengths), searches for the additional Higgs bosons that come within the model, both direct and indirect, and, finally, theory considerations based on perturbativity, unitarity, triviality and vacuum stability. A statistical analysis is then performed, with the likelihood function expressing the plausibilities of different parameter values for the given samples of data. The 2HDM parameter space is 6-dimensional (after enforcing $m_h$ reconstruction), so the standard way of extracting bounds is projecting the full parameter space onto 2-dimensional planes, defined by any two model parameters. The statistical procedure is typically to maximise the (log) likelihood of the four remaining parameters.

\section{Parameter Space}

\noindent
Here, we intend to detail how both theoretical and experimental constraints were applied onto the 2HDM type-I and -II parameter spaces and how cross-sections were computed over those as well as how \Magellan was used for this purpose. 

\subsection{Tools}
As we have shown, there are ultimately 6 inputs that make up the 2HDM parameter space. In the basis we are working in, these are   $m_H$, $m_{H^\pm}$, $m_A$, $\cos(\beta - \alpha)$, $\tan\beta$ and $Z_7$. It is possible that the allowed regions of this parameter space are not all interconnected,
instead being made up of a series of pockets of valid points. A straightforward grid-type scan was found to be highly inefficient in such a case and thus was abandoned early on. In order to maximise the efficiency of our parameter space scans we used the software package \Magellan. This combines multiple  tools (\HiggsBounds\cite{HB}, \HiggsSignals\cite{Bechtle_2021}\cite{HS1}\cite{HS2}, \THDMC\cite{2HDMC} and \TTPS\cite{T3PS}) into a single pipeline, allowing  users to generate points in the allowed parameter space of a given 2HDM. The points are found more efficiently thanks to \TTPS, a Markov Chain Monte Carlo (MCMC) generator\cite{T3PS}, with packages \HiggsBounds, \HiggsSignals and \THDMC contributing to the likelihood used by the MCMC and checking that points are valid before being output for phenomenological use. Following this, \Magellan invokes directly  \MadGraph \cite{MG2014}, which then calculates the cross-sections for the three contributing sub-processes individually. This was done for both the type-I and -II realisations of the 2HDM. For type-I we used the model file 'THDM\_type1\_UFO', created  using \FeynRules \cite{Feyn2012,Feyn2014} while  for type-II we used the readily available '2HDMtII{\textunderscore}NLO' \cite{typeII2018}, both of which we used at Leading Order (LO). Finally, we used a default Parton Distribution Function (PDF) set that comes with \MadGraph, the NNPDF 2.3 one \cite{Deans:2013mha}, also taken at LO  for consistency\footnote{}\footnotetext[4]{We are interested here in relative effects between the 2HDM and the SM in our three processes of reference, for which QCD corrections are essentially the same.}.

\subsection{Constraints}

\begin{figure}[H]
	\begin{center}
		\includegraphics[width=0.61\linewidth]{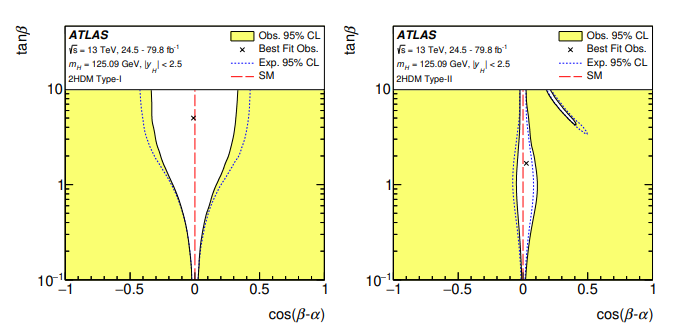}
		\caption{\label{fig:ATLAS:cba_tb}
			Allowed regions for the \cba and \tb parameters in the 2HDM models type-I and -II, on the left and right, respectively, for observations made by ATLAS. These are obtained assuming that the $125$ GeV boson is the light, CP-even Higgs boson, $h$, of the 2HDM. Constraints are seen to be tighter on type-II than on type-I . Plots are taken directly from \cite{ATLAS2020}.
		}
	\end{center}
\end{figure}

\begin{figure}[!htb]
	\begin{center}
		\includegraphics[width=0.61\linewidth]{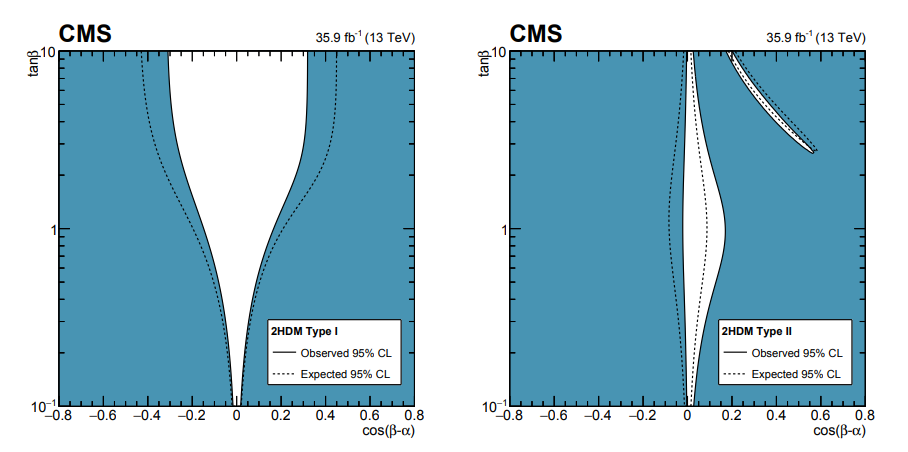}
		\caption{\label{fig:CMS:cba_tb}
			Allowed regions for the \cba and \tb parameters in the 2HDM models type-I and -II, on the left and right, respectively, for observations made by CMS. These are obtained assuming that the $125$ GeV boson is the light, CP-even Higgs boson, $h$, of the 2HDM. Constraints are seen to be tighter on type-II than on type-I  Plots are taken directly from \cite{CMS2019}.
		}
	\end{center}
\end{figure}

In our analysis,  theoretical constraints were imported from \THDMC. On the experimental side, constraints were included from multiple sources.  Firstly, \HiggsBounds uses the limits found from Higgs searches at the LHC, Tevatron and LEP to determine if a given set of Higgs sector predictions, for each 2HDM considered,  have been excluded at a 95\% Confidence Level (CL) or otherwise.    Both ATLAS and CMS have done analyses on the parameter space of the 2HDM based on their observations at the LHC. From ATLAS, in Fig.~\ref{fig:ATLAS:cba_tb}, we can clearly see that the constraints for the type-II case are much tighter than those for type-I, thus there is a larger parameter space to be scanned for potentially high cross-sections. We see a very similar picture when we look at equivalent plots from CMS in Fig.~\ref{fig:CMS:cba_tb}. (Our results based on \Magellan are very similar.)
Secondly, \HiggsSignals is exploited to check the consistency of the measured properties of the discovered Higgs boson with those of the lightest CP-even Higgs state of the 2HDHM, $h$, by performing a $\chi^2$ analysis. In fact, 
\HiggsSignals compares the predictions for 2HDM scalar sectors to the signal rates and mass measurements of the SM-like Higgs boson at the LHC, so this allows for a likelihood estimate to be calculated.
However, given the fact that the $h$  state, possibly alongside $H$, $A$ and $H^\pm$, also enters EW Precision Observables  
(EWPOs), a proper $\chi^2$ analysis ought to be performed across the two datasets, which is what \Magellan does. While \THDMC also computes the EWPO constraints, those used in this analysis are taken from Ref. ~\cite{Gfitter} and used in conjunction with those  calculated by \HiggsSignals. Specifically, 
an overall $\chi^2$ value is determined by taking the one given by \HiggsSignals and calculating one using the values of best fit in \Gfitter for $S$ and $T$ where $U$ is set to 0.  In order to combine these  we use the following expressions:
\begin{equation}
	\chi^{2}_{\rm tot} = \chi^{2}_{HS} + \chi^{2}_{ST},
	\label{chitot}
\end{equation}
\noindent
where $\chi^{2}_{HS}$ is the $\chi^2$ for \HiggsSignals and $\chi^{2}_{ST}$ is the one for an $S$ and $T$ compatibility measure given by
\begin{equation}
	\chi^{2}_{ST} = \frac{(S - S^{\rm exp}_{\rm best\, fit})^2}{\sigma^{2}_S (1- \rho^{2}_{ST})} + \frac{(T - T^{\rm exp}_{\rm best\, fit})^2}{\sigma^{2}_T (1- \rho^{2}_{ST})} -2\rho_{ST} \frac{(S - S^{\rm exp}_{\rm best\, fit})(T - T^{\rm exp}_{\rm best\, fit})^2}{\sigma_T \sigma_S (1- \rho^{2}_{ST})}.
\end{equation}
It is $\chi^2_{\rm tot}$ that we test then, as explained in \cite{Accomando:2022nfc}.

Of all possible solutions to these theoretical and experimental constraints, there are some corresponding to the
wrong-sign solution that we described already. It turns out that these play a crucial role in the phenomenology of  SM-like Higgs boson production in association with single-top at the LHC. It has been shown that, in order to obtain a wrong-sign solution in the 2HDM, one has to require that $\sin(\beta - \alpha)>0$ \cite{rdcdT2}, as the available parameter space for  negative values has essentially been ruled out. Furthermore, a lower bound for the charged Higgs boson mass in the type-II was found to be $m_{H\pm} > 580$ GeV  in 
Ref.~\cite{chargemass},  in turn rendering the cross-section contributions of $H^\pm$ propagators (replacing
the $W^\pm$ ones in Fig.~\ref{Feyn_1}) negligible, so that we have altogether neglected these in the calculation. 
Such a severe bound on $m_{H^\pm}$ does not strictly apply to the 2HDM type-I, however, for consistency, we have removed the corresponding Feynman diagrams in the corresponding cross-section calculations (by making the $H^\pm$ state heavy enough). Finally, 
a
 general limit on 2HDMs with a (softly-broken) $Z_{2}$ symmetry is given in Ref. ~\cite{Sanyal_2019} as $\tan\beta \geq 1$, this is the constraint used here on the type-I case. For the type-II one, there is a stronger restriction on $\tan\beta$ from the mass of the charged Higgs boson itself, as was shown in Ref.~\cite{chargemass} (c.f. Fig.~4 therein). Herein, the lower limit for the type-II scenario was then chosen as $\tan\beta\geq 5$. 

\subsection{Cross-sections at the LHC}

\begin{figure}[htb!]
	\begin{center}
		\includegraphics[width=0.385\linewidth]{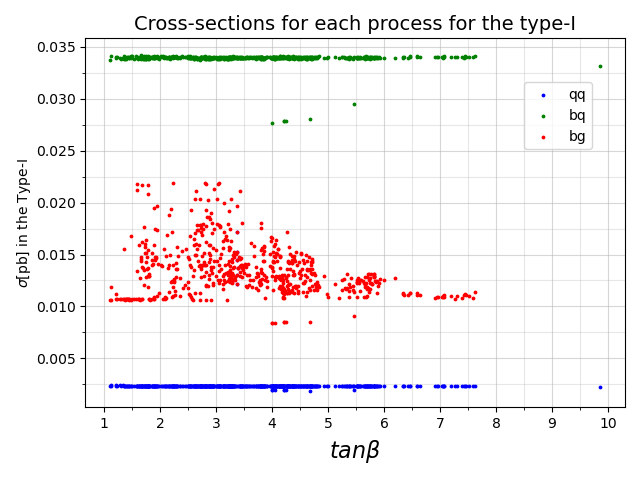}
		\includegraphics[width=0.405\linewidth]{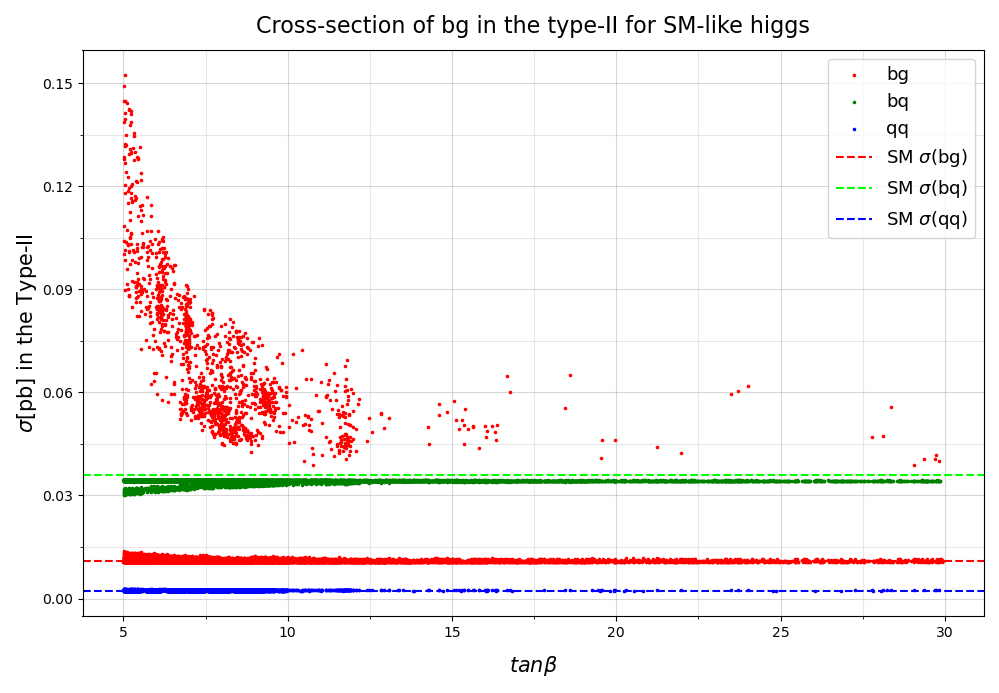}
		\caption{\label{tancross}
			Cross-sections of points obtained in our described scans over the parameter space for the 2HDM type-I (left) and -II  (right) plotted against the value of $\tan\beta$. (Note that these two plots are {not} to the same scale as the highest cross-sections in type-II are considerably larger than in type-I.)}
	\end{center}
\end{figure}

Fig.~\ref{tancross} shows the initial output by \MadGraph for representative points passing the aforementioned theoretical and experimental constraints, split across the three sub-processes described previously: bq, bg and qq. Herein, the underlying horizontal 
lines represent the cross-sections for these channels as obtained in the SM.
We can see that the type-I points appear to behave much like the SM predicts, with some points even having a lower cross-section. While there is a slight increase in the {bg} process, this does not seem to show any significant excess which would make it  distinguishable from the SM at the LHC. The type-II, however, looks very different. While the bq and qq sub-processes for the type-II do behave like the SM, showing cross-sections similar to the SM ones and the expected hierarchy between themselves, the bg sub-process does not adhere to this pattern. Instead, we see that the latter  becomes dominant over the leading one in the SM (bq): in some places, this is by a very large margin, in fact. The highest point in the type-II distribution has a cross-section for bg that is over 5 times  the size of the bq one, presenting us with the possibility of promptly extracting the bg channel at the LHC as well as ascribing it to the 2HDM type-II hypothesis. 

\begin{figure}[H]
	\begin{center}
		\includegraphics[width=0.49\linewidth]{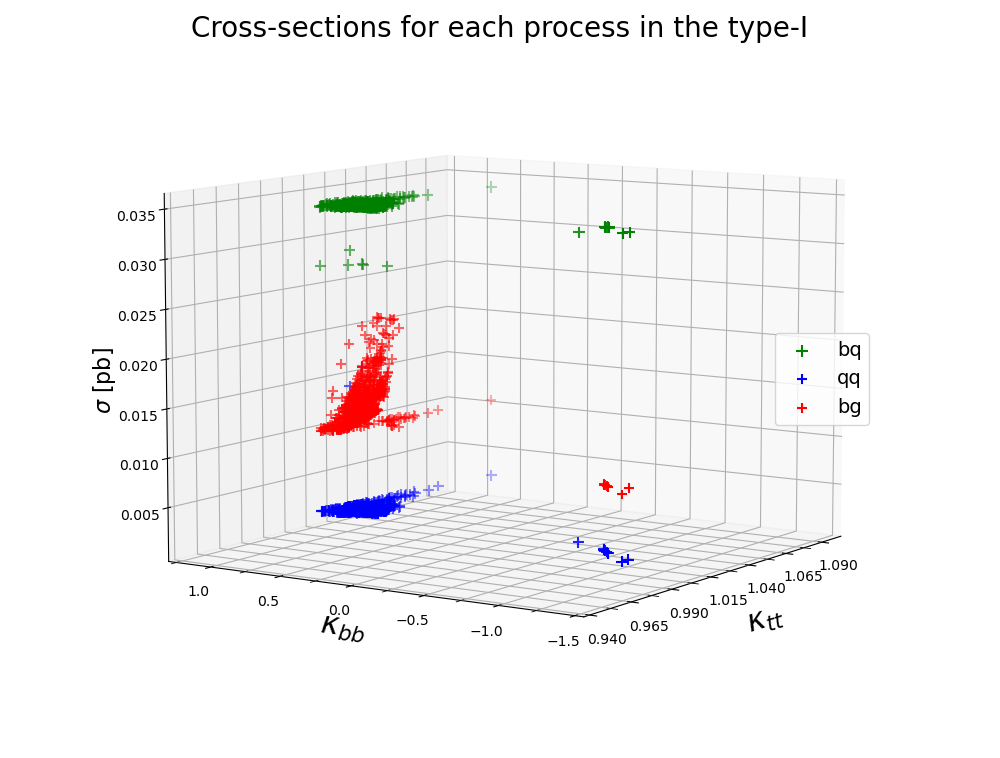}
		\includegraphics[width=0.39\linewidth]{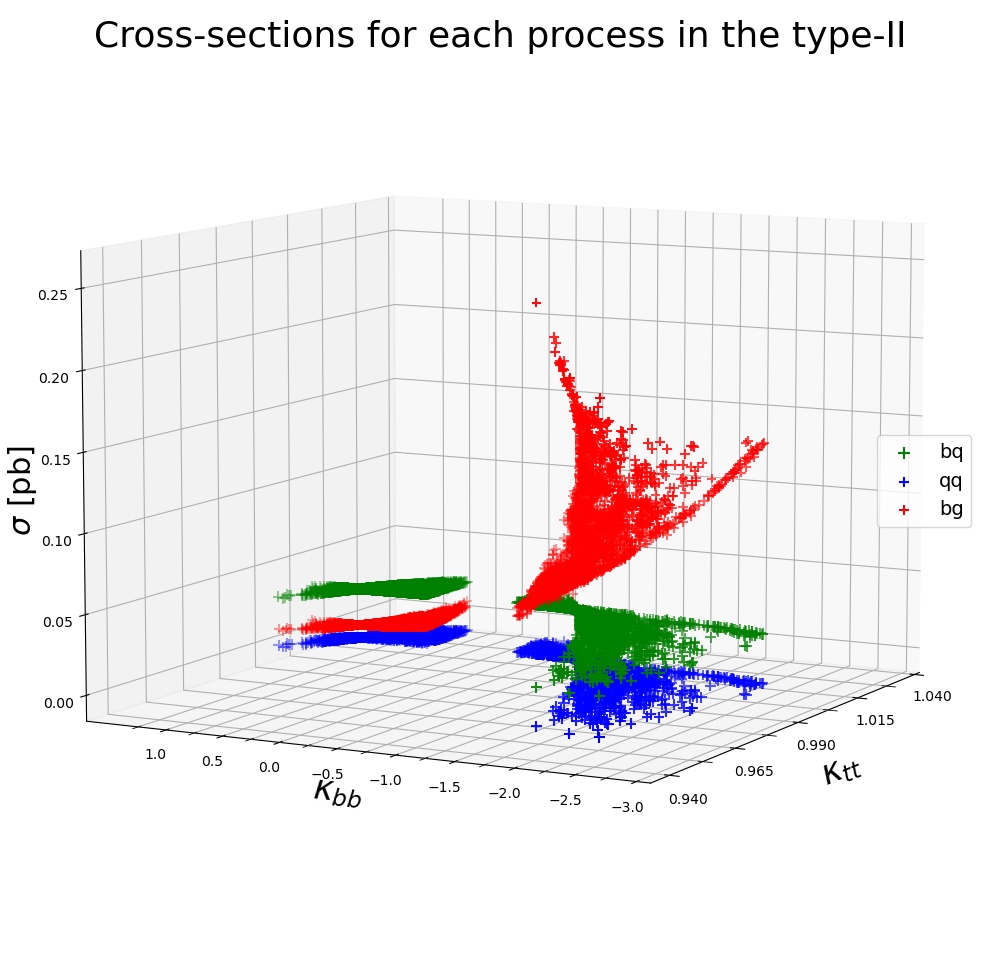}
		\caption{\label{3D_kappas}
			Cross-sections of points obtained in our described scans over the parameter space for the 2HDM type-I (left) and -II  (right) plotted against $\kappa_{bb}$ and $\kappa_{tt}$. Note that the $\kappa_{bb}$,$\kappa_{tt}$ planes are tilted to provide a better view of the points in the 3D space, in particular the highest point of the type-II plot has a cross-section of $\approx 0.25$pb (Again, note that these two plots are \textbf{not} to the same scale as the highest cross-sections in type-II are considerably larger than in type-I.)}
	\end{center}
\end{figure}

Looking now at Fig.~\ref{3D_kappas}, where we plot the cross-sections versus the (rescaled)  Yukawas entering the three sub-processes,  
we can establish that the points found for the type-I do not display a significant variation in cross-section size between the alignment limit and wrong-sign scenario for the bq and qq sub-processes. There is instead a clear boost in the {bg} cross-section, but this occurs in the alignment limit only  and is not as significant as the boost we have seen in the type-II. (Therefore, henceforth, we will drop consideration of this case.) Turning to the plot for the type-II case, for the {bg} process,  it  appears that both wrong-sign and alignment points offer the discussed increase   in  cross-section. Again, for both the bq and qq  sub-processes, also in type-II, the two  regions of parameter space are similar in terms of (mild) variations in cross-section with respect to the SM.   



\begin{figure}[H]
	\begin{center}
		\includegraphics[width=0.425\linewidth]{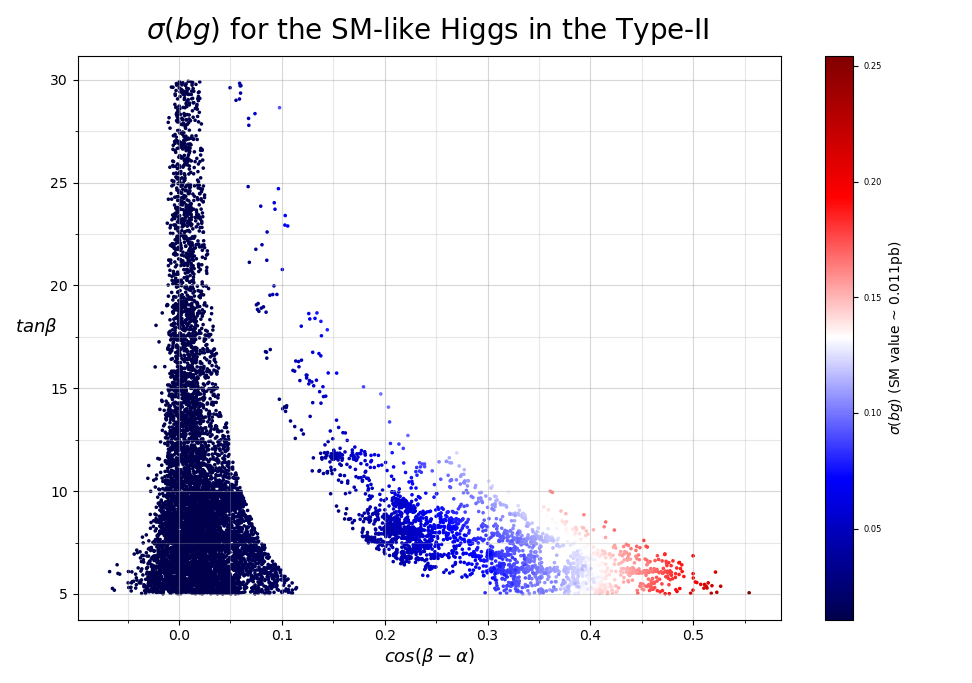}
		\includegraphics[width=0.4\linewidth]{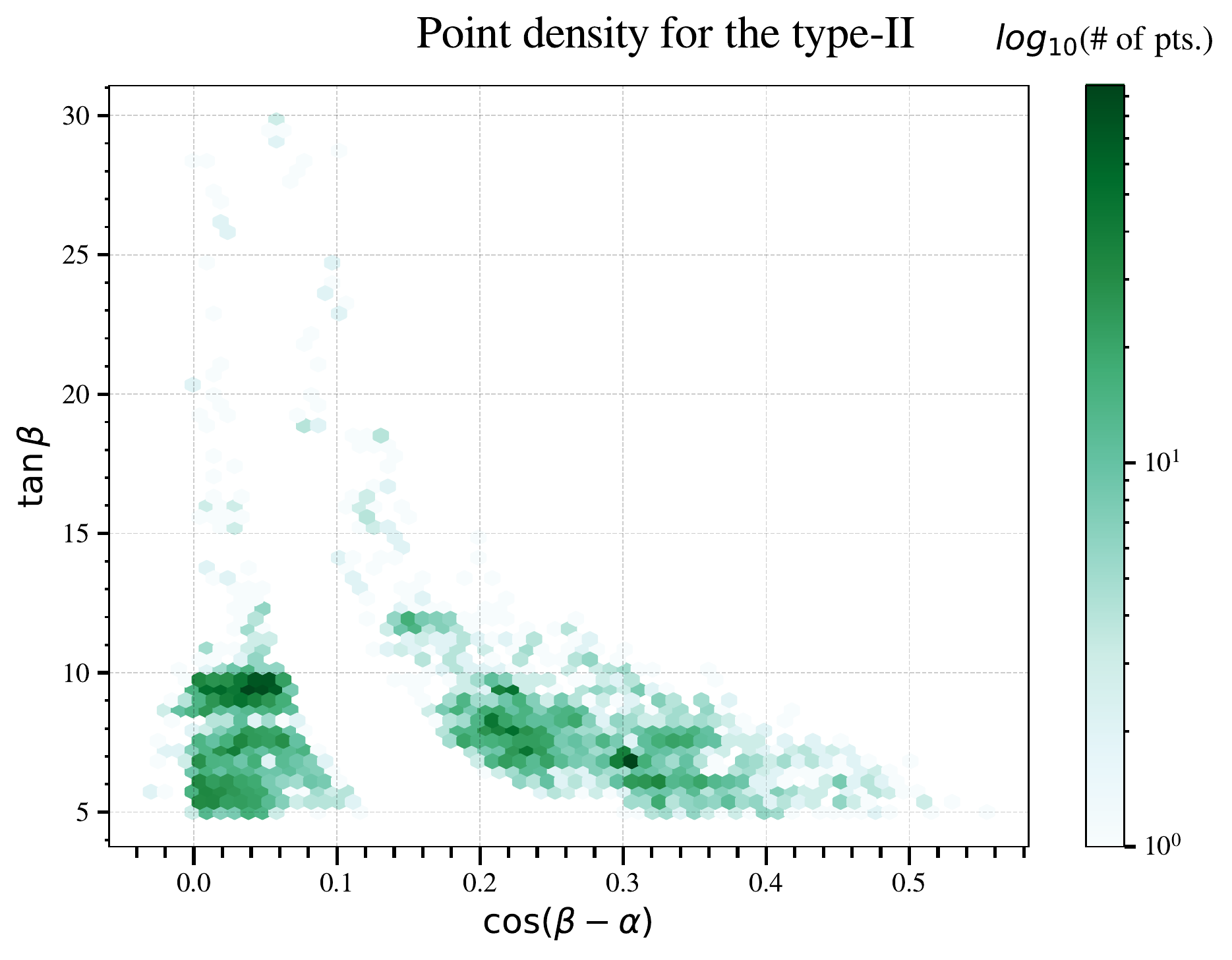}
		\caption{\label{costan}
                                Cross-sections (left) and $\log_{10}$ of the number (right)  of points obtained in our described scan of type-II
for the bg sub-processes mapped over the ($\cos(\beta -\alpha)$, $\tan\beta$) plane. (Recall that the SM cross-sections is 0.011 pb.)  
		}
	\end{center}
\end{figure}

In Fig.~\ref{costan}, we examine the cross-section for the bg sub-process over the $(\cos(\beta - \alpha)$, $\tan\beta$) plane. In the left  plot the colour of a point indicates how large the cross-section found for it is, with the associated colour scale on the right. It is important to remember here that the SM value is $\approx$ 0.011 pb, i.e., only the very darkest points correspond to this value, the vast majority being of greater magnitude, with the highest ones being more concentrated in the right-arm area. Turning to the right plot, here, the colour  indicates the number of points that have been binned into each coloured hexagon. The colour scale is logarithmic and shown on the right of the plot. We see that medium values of $\tan\beta$ are favoured, in particular, between 5 and 10. There is also a very concentrated region of points around $\tan\beta\approx9$ and for $0 \leq \cos(\beta - \alpha) \leq 0.07$, which indeed captures large cross-sections values too.

\begin{figure}[H]
	\begin{center}
		\includegraphics[width=0.6\linewidth]{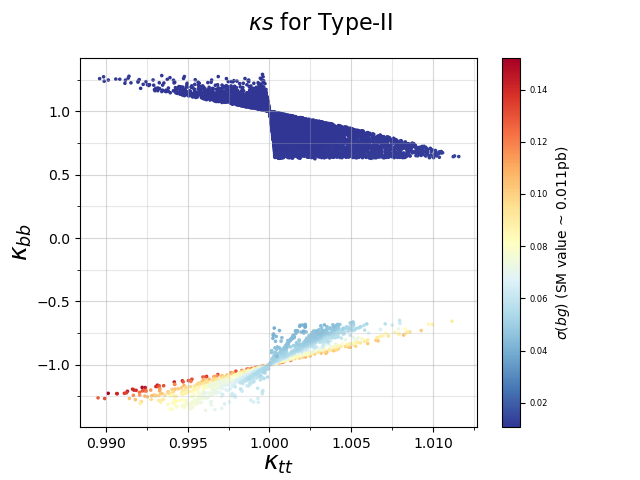}
		\caption{\label{kbbktt}
			Cross-section of points obtained in our described scan of type-II 
for the bg sub-processes mapped over the  ($\kappa_{tt}$, $\kappa_{bb}$) plane. Here we display only the points with a 'good' ${\chi_{Tot}}^2$ value. (Recall that the SM cross-sections is 0.011 pb.) 
		}
	\end{center}
\end{figure}

In Fig~.\ref{kbbktt}, we project the points over the plane of $\kappa_{tt}$ and $\kappa_{bb}$, limitedly to type-II. Again,  we use a colour gradient to indicate the cross-section of each point. In the left plot we can see that the vast majority of points are SM-like in $\kappa_{tt}$, though there is a small cluster of  `doubly wrong-sign solutions'  on the left  of the plot. (Here, doubly wrong-sign refers to points where  $\kappa_{bb}$ and $\kappa_{tt}$ have both opposite sign with respect to the SM.) The highest cross-section points are found when $\kappa_{tt}$ is SM-like, though, so we zoom in on this region in the right frame of the figure. Altogether, something that has now  become clear is that, while $\kappa_{tt}$ is constrained into quite small regions, $\kappa_{tt}\approx -1$ and $0.94 \leq \kappa_{tt} \lessapprox1$,  the same cannot be said of $\kappa_{bb}$, for which the region $-3 \leq \kappa_{bb} \leq 1.5$ is accessible, with the highest cross-sections found for the alignment points around $0.75$ and for  the wrong-sign ones around $-1$. 
\begin{figure}[H]
	\centering
	\begin{subfigure}[b]{.6\linewidth}
		\includegraphics[width=\linewidth]{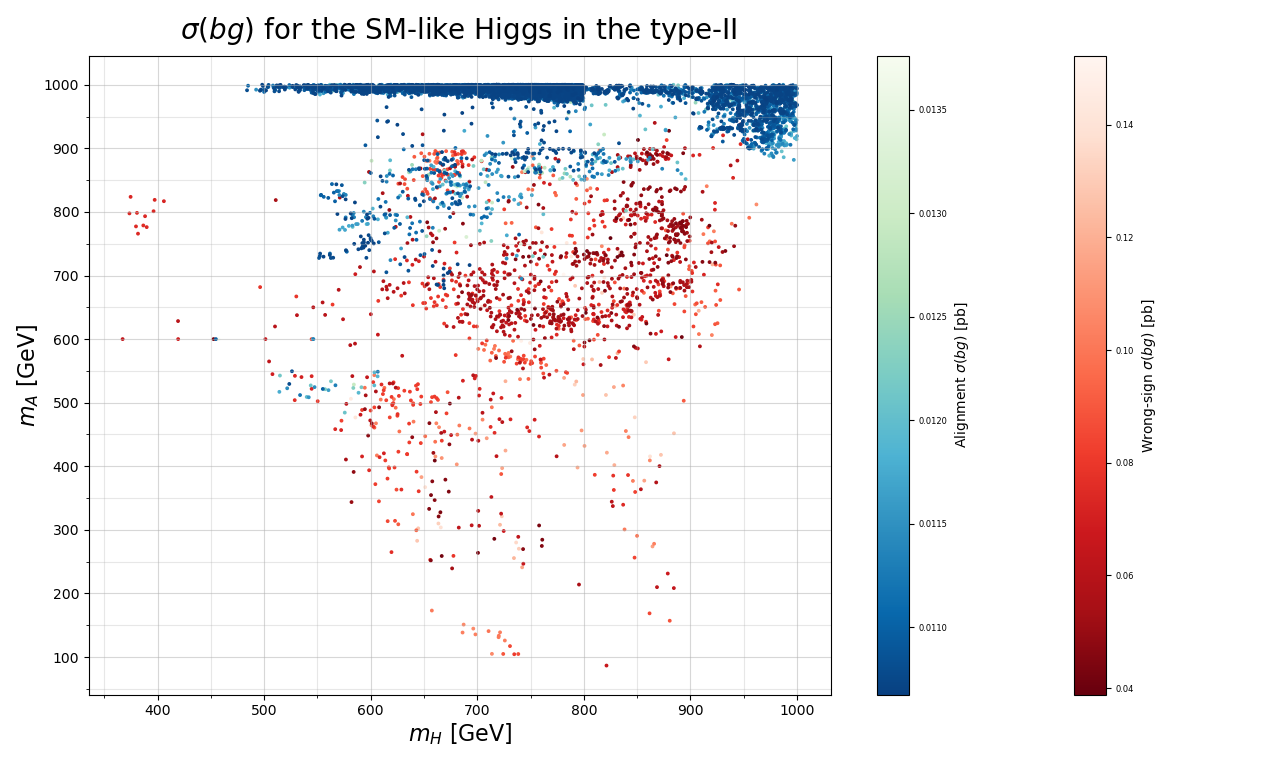}
	\end{subfigure}
	\begin{subfigure}[b]{.44\linewidth}
		\includegraphics[width=\linewidth]{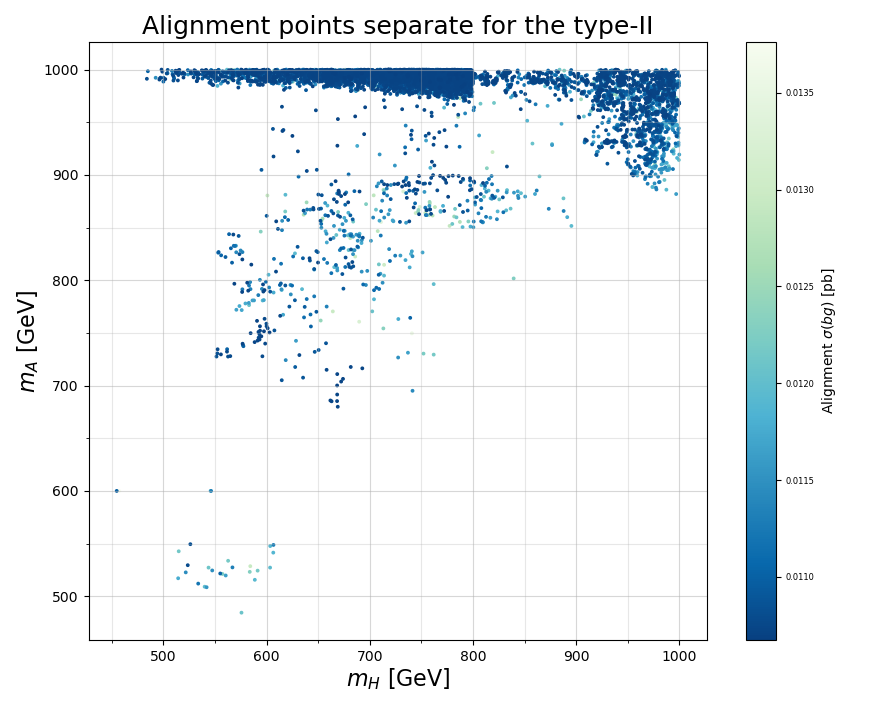}
	\end{subfigure}
	\begin{subfigure}[b]{.46\linewidth}
		\includegraphics[width=\linewidth]{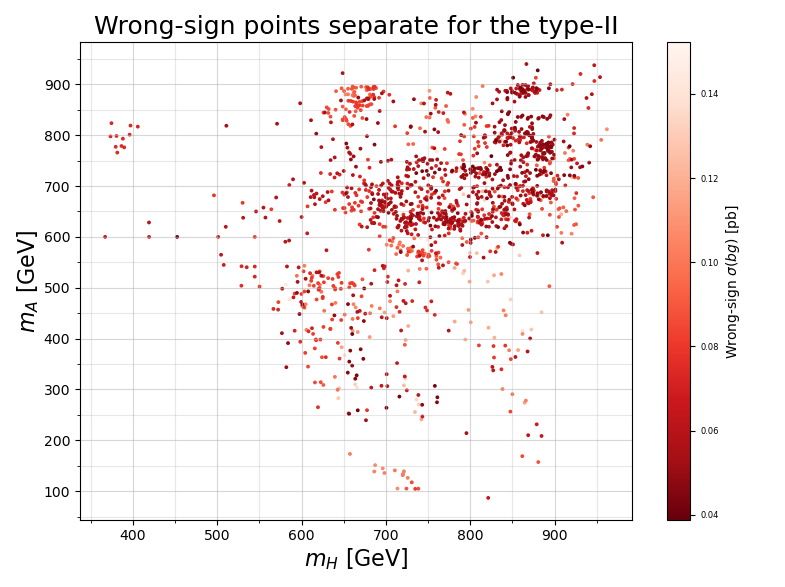}
	\end{subfigure}
	\caption{\label{masses}
		Cross-section of points obtained in our described scan of type-II
for the bg sub-processes mapped over the ($m_H$, $m_A$) plane. The top plot is split into two colour palettes, the blue-green one  is only alignment points while the red-orange one is only wrong-sign points. Both sets are coloured in a gradient shown in their respective colour bars. The lower plots are  coloured according to the size of the bg cross-section, as indicated by the associated colour bars.
	}
\end{figure}

In Fig.~\ref{masses},  we examine the ($m_A$, $m_H$) plane, again, limitedly to type-II. The bottom-left frame shows a large number of the highest bg  cross-sections clustered in a region of high mass for both particles, although it also shows surprisingly low bg cross-sections for many points with a high mass for the $A$ state  and medium-to-high mass for the $H$ state. The bottom-right plot helps us to make some sense of this: these  points are all from the alignment limit. (The top frame combines the two regions.) In short, there seems to be a great deal more variation in the bg cross-sections of the alignment limit points than for the wrong-sign ones, which consistently produce much higher than SM cross-sections. Notice that neither the $H$ nor the $A$ state enters the cross-sections of our sub-processes, so that this figure is presented for the sole purpose of completing the full mapping of the phenomenologically interesting region of (type-II) parameter space for the bg channel. (As for the $H^\pm$ state, we simply reiterate here that its mass is constrained to be above 580 GeV or so and has no direct implications for the cross-section of the bg process in the two parameter space regions of reference.)

\section{Analysis}
The purpose of this part of our study is to determine the possibility of detecting 2HDM type-II cross-sections at the HL-LHC for the production of $h$ in association with a single-top and a $W^\pm$ via the bg sub-process for the `wrong-sign solution' of the bottom (anti)quark Yukawa coupling. To this end, we perform a realistic MC analysis  with $h$ decaying into $b \bar b $ pairs to prove that this phenomenology would clearly be manifest at the detector level. Specifically, we test whether the signal emerging from the 2HDM type-II scenario in such a parameter configuration might lead to any differences in differential 
distributions  (e.g., invariant and/or transverse masses, transverse momenta, etc.) 
 for the final state particles when compared to the SM case. We approach the topic by considering the main backgrounds to eventually, after a dedicated cutflow,  calculate signal-to-background significances to attest the above-mentioned statement. 

The plan of the section is as follows.  We describe the MC analysis that we will perform (i.e., simulation tools, cutflow, etc.). After which we will present our parton and hadron level results. Finally, in the last sub-section, we will evaluate signal-to-background 
significances in the comparison between 2HDM and SM yields.
\subsection{Methodology}
In this sub-section, we provide the simulation details and cutflow information used to conduct our analysis.
\subsubsection{Simulation Details}
We generate samples of events with $\sqrt{s}$ = 13.6 TeV for the LHC energy. Our study is based on an integrated luminosity of $3000$ fb$^{-1}$, an integrated luminosity expected to be attainable at the HL-LHC. First, we generate events using the SM 
implementation of the bg sub-process. Second, we consider a sample Benchmark Point (BP) in the 2HDM type-II framework where, as intimated,  we fix the lighter CP-even Higgs boson to be the SM-like Higgs boson with $m_h=125$ GeV. This BP has been tested against theoretical and experimental constraints by using \HiggsBounds \cite{HB}, \HiggsSignals \cite{Bechtle_2021,HS1,HS2} and \THDMC \cite{2HDMC}. Our study is aimed at events that contains $h\rightarrow b \bar b$ decays with the top (anti)quark decaying into leptons plus $\ b$-jet while the (primary) $W^\pm$ boson in the bg sub-process decays leptonically. The production cross-section at the Leading Order (LO) and the 2HDM type-II input parameters for the full decay chain of the process are shown in Tab.~\ref{tab:params}. The corresponding SM cross-section value at LO level is 0.000187 pb. For our study, we have used the ${\tt NNPDF23_{\bf -}lo_{\bf -}as_{\bf -}0130_{\bf -}qed}$ \cite{Deans:2013mha} set  to model the Parton Distribution Function (PDF) (with default settings). 
\begin{table}[htb!]
\begin{center}
\hspace*{-1.75cm}
\scalebox{1.3}{
\begin{tabular}{ |c|c|c|c|c|c|c|c|c| }
 \hline
 Label & $m_h$ (GeV) & $m_H$ (GeV) & $\tan\beta$ & $\sin (\beta -\alpha)$ & $m_{12}^2$ (GeV$^2$) & BR($h\rightarrow b \bar b$) &$\sigma$(pb)  \\
 \hline
BP & 125 & 781 & 5.06049& 0.906849 & 113436 &  6.85754$\times 10^{-1}$ &0.00244\\
 \hline
\end{tabular}
}
\caption{\label{tab:params} The 2HDM parameters and cross-section at LO of the bg process with $h \rightarrow b \bar b$  decay channel for the 2HDM type-II BP.}
\end{center}
\end{table}

\vspace*{1em}
\tikzstyle{node} = [rectangle, rounded corners, minimum width=3cm, minimum height=1cm,text centered, draw=black]
\tikzstyle{arrow} = [thick,->,>=stealth]

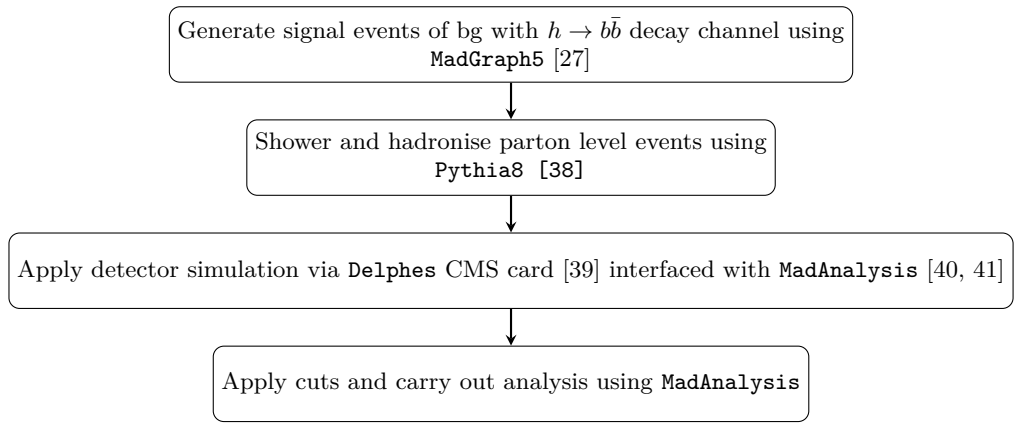
\begin{figure}[H]
	\centering
	\hspace*{-0.5truecm}
	\begin{tikzpicture}[node distance=1.5cm]
		\node (step1) [node, align=center] {Generate signal events of bg with $h \rightarrow b \bar b$ decay channel using \\ \MadGraph \cite{MG2014}};
		\node (step2) [node, align=center, below of=step1] {Shower and hadronise parton level events using\\ \tt{Pythia8} \cite{SJOSTRAND2008852}};
		\node (step3) [node, align=center, below of=step2] {Apply detector simulation via \Delph CMS card \cite{deFavereau:2013fsa} interfaced with \MadAnalysis \cite{Conte2013, conte2018}};
		\node (step4) [node, align=center, below of=step3] {Apply cuts and carry out analysis using \MadAnalysis};
		
		\draw [arrow] (step1) -- (step2);
		\draw [arrow] (step2) -- (step3);
		\draw [arrow] (step3) -- (step4);
	\end{tikzpicture}
	\caption{Illustration of the procedure used to generate and analyse MC events.}
	\label{fig:toolbox}
\end{figure}

The procedure illustrated in Fig.~\ref{fig:toolbox} was used to generate and analyse signal events to carry out our realistic MC simulation. The same procedure was used to generate background events too. For the latter, we considered the following SM processes: $gg,q\bar q \to t \bar t$, $gg,q\bar q \to t \bar t h$, $gg,q\bar q \to t \bar t b \bar b$, $gg,q\bar q \to t \bar t t \bar t$, $q\bar q \to W^+W^- h$, $q\bar q \to ZZh$, and  $q\bar q \to ZW^+W^-$. 
\subsubsection{Cutflow}
In this sub-section, we describe the cuts used to identify the signal and reduce its backgrounds. Electrons are reconstructed and required to satisfy $p_T > 7$ GeV and $|\eta| < 2.5$ whereas reconstructed muons are required to satisfy $p_T > 5$ GeV and $|\eta| < 2.4$. The jets are clustered using the anti-$k_T$ algorithm \cite{Cacciari_2008} with fixed cone size, $R=0.4$. The jets considered in the analysis are then required to satisfy $p_T > 25$ GeV and $|\eta| < 5.0$. Jets reconstructed within $|\eta| < 2.4$ are identified as $b$-jets, in presence of $b$-tagging. Concerning the latter, 
in this paper,  we adopt and implement simplified tagging and mis-tagging procedures. A loose $b$-jet selection efficiency of 84$\%$ is applied. The respective mis-tagging rates for gluon jets ($c$-jets) and light-quark jets are 1.1$\%$ and 11$\%$. The mis-tagging rates are a bit counter intuitive with the $c$-jet rate 10 times smaller than the light-quark rate. This is mainly because the tagging does a good job at rejecting $c$-jets and the high rate of light quarks. In fact, we use cuts that are within trigger efficiencies and $b$-tagging rates as measured by the CMS experiment \cite{CMS-PAS-HIG-19-008}.  More details about cutflow analysis can be found in Fig.~\ref{fig:cutflow}.
\vspace*{1em}
\tikzstyle{node} = [rectangle, rounded corners, minimum width=3cm, minimum height=1cm,text centered, draw=black]
\tikzstyle{arrow} = [thick,->,>=stealth]

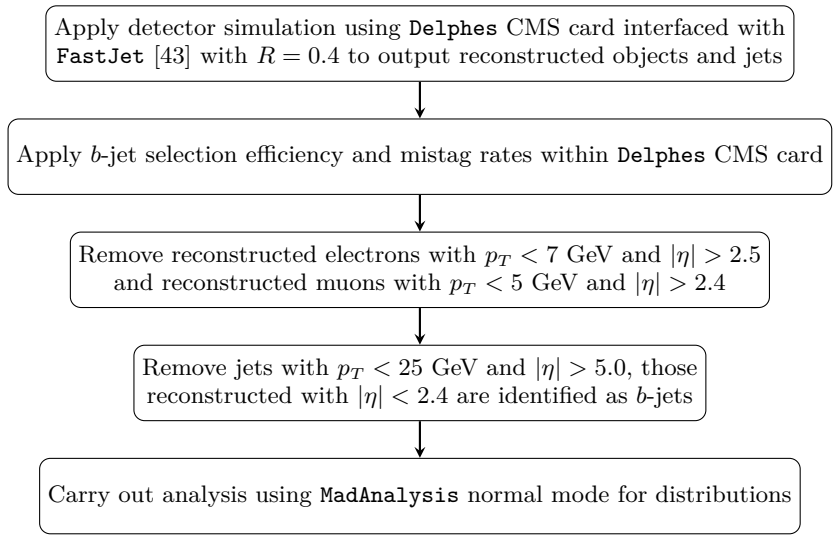
\begin{figure}[H]
	\centering
	
	\begin{tikzpicture}[node distance=1.5cm]
		\node (cut1) [node, align=center] {Apply detector simulation using \Delph CMS card  interfaced with\\\fast \cite{Cacciari:2011ma} with $R=0.4$ to output reconstructed objects and jets};
		\node (cut2) [node, align=center, below of=cut1] {Apply $b$-jet selection efficiency and mistag rates within  \Delph CMS card};
		\node (cut3) [node, align=center, below of=cut2] {Remove reconstructed  electrons with $p_T < 7$ GeV and $|\eta| > 2.5$ \\ and reconstructed muons with $p_T < 5$ GeV and $|\eta| > 2.4$};
		\node (cut4) [node, align=center, below of=cut3] {Remove jets with $p_T < 25$ GeV and $|\eta| > 5.0$, those \\ reconstructed with $|\eta| < 2.4$ are identified as $b$-jets};
		\node (cut5) [node, align=center, below of=cut4] {Carry out analysis using \MadAnalysis  normal mode for distributions};
		
		\draw [arrow] (cut1) -- (cut2);
		\draw [arrow] (cut2) -- (cut3);
		\draw [arrow] (cut3) -- (cut4);
		\draw [arrow] (cut4) -- (cut5);
	\end{tikzpicture}
	\caption{Illustration of the initial procedure for event reconstruction and jet clustering.}
	\label{fig:cutflow}
\end{figure}
\subsection{Results}
In this sub-section, we present our results at parton and hadron level, in turn. We then also present a signal-to-background analysis at the detector level, finally computing (after the implementation of the described cutflow) the significances for the chosen 2HDM type-II BP and compare these to those attainable in the SM.

\subsubsection{Parton Level Analysis}
In this sub-section, we perform a parton level analysis of events emerging from the 2HDM type-II and SM for the purpose of identifying differences in differential distributions that may eventually be pursued at hadron level to characterise the signal as being of BSM origin, above and beyond the fact that the integrated cross-section yields of the two theoretical scenarios are very different.
Hence, we will be looking at (identically) normalised distributions to extract the shapes only. 

\begin{figure}[H]
	\begin{center}
		\includegraphics[width=0.45\linewidth]{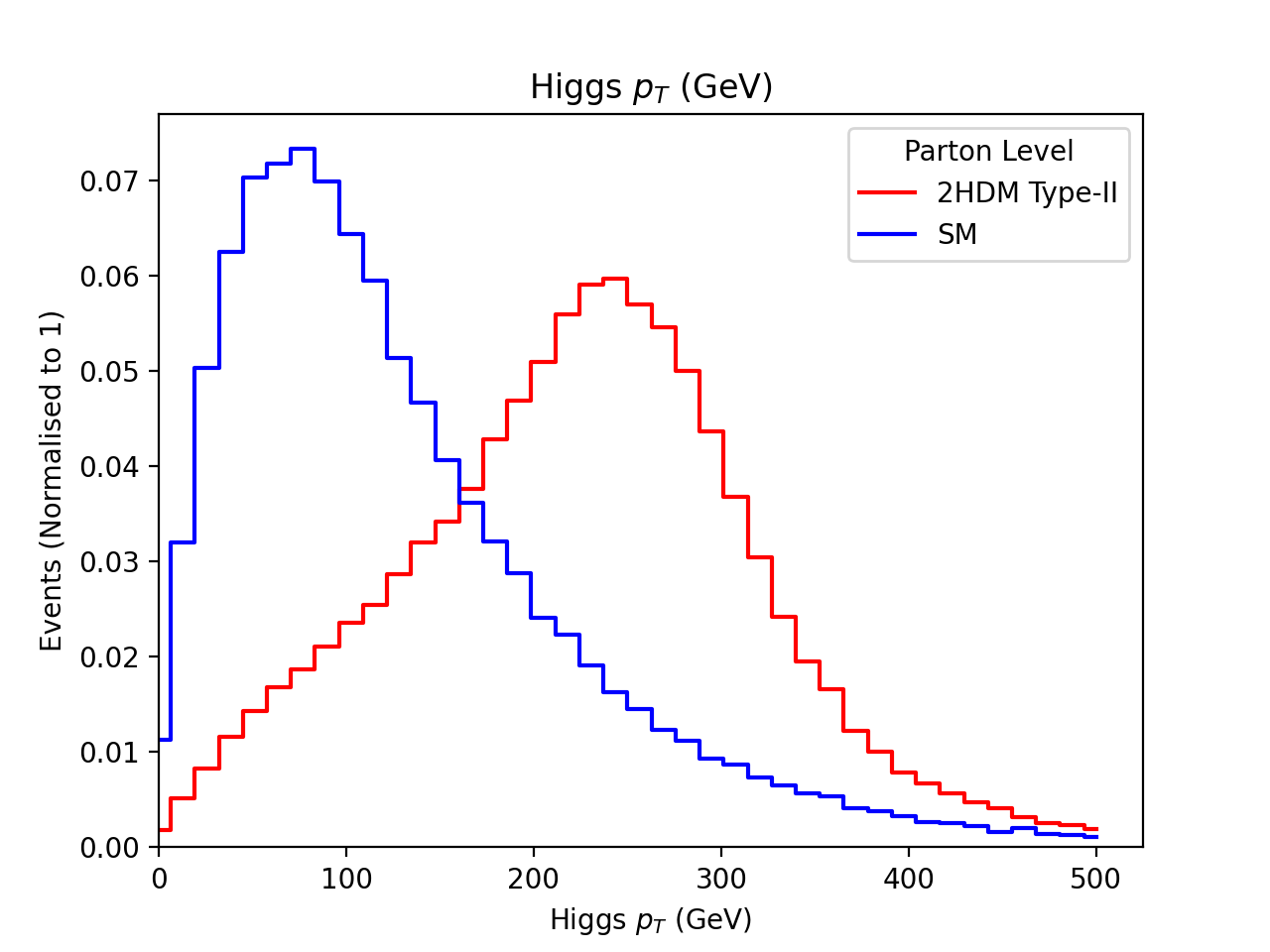}
		\includegraphics[width=0.45\linewidth]{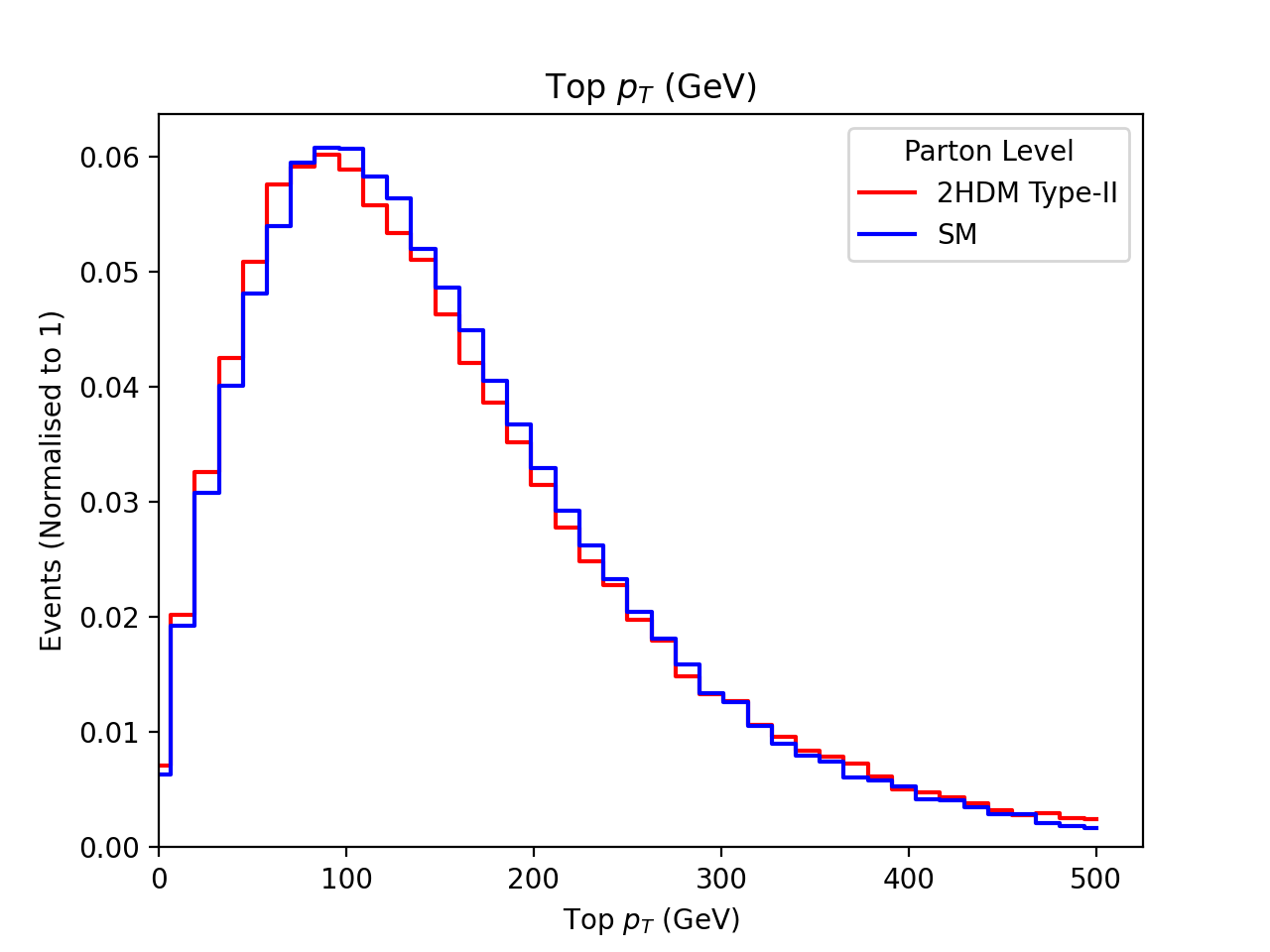}
		\\
		\includegraphics[width=0.45\linewidth]{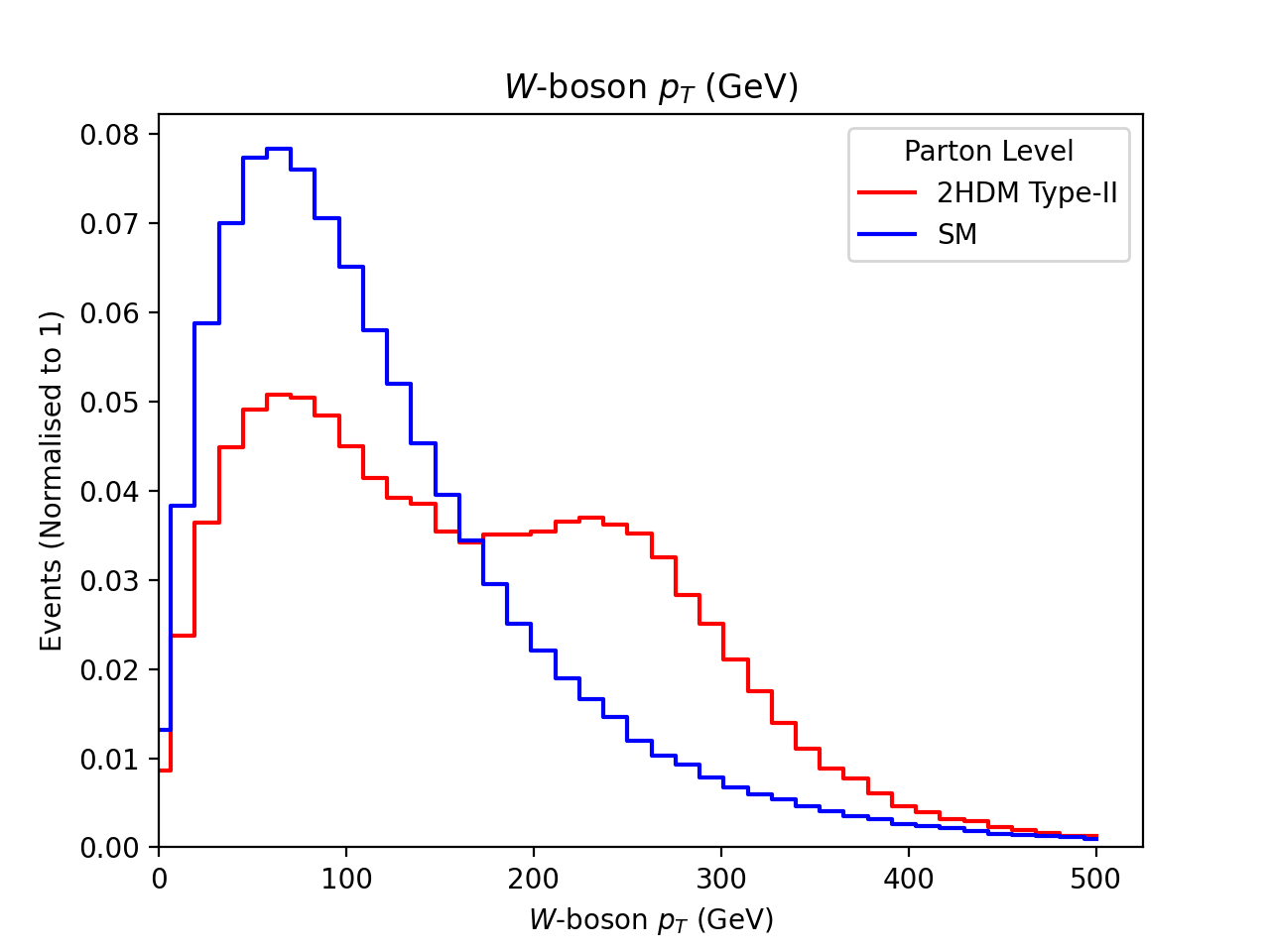}

		\caption{\label{parton_level}
			Upper panel: Normalised transverse momentum distributions of the Higgs boson (left) and top (anti)quark (right) at the parton level. Lower panel: Normalised transverse momenta distributions of the $W^\pm$ boson at the parton level.  
		}
	\end{center}
\end{figure}

We show in Fig.~\ref{parton_level}  the $p_T$ distributions for the 2HDM type-II and SM events for the three heavy objects in the final state of the bg sub-process: the Higgs boson, top (anti)quark and charged gauge vector. Herein,  one can see that, for the $h$ and $W^\pm$ states, there is a stark difference between the $p_T$ distributions in the two theoretical scenarios:  in the 2HDM type-II they are much harder than  in  the SM due to the different relative kinematics of the two models, in turn driven by the different $\kappa$ values entering the cross-section (notably, $\kappa_b$). In contrast, the $p_T$ distribution for the top (anti)quark remains more or less the same in the 2HDM type-II when compared to the SM. Thus, assuming an efficient reconstruction at the detector level of the $h$ boson (notice that the top (anti)quark decay would also produce a $b$-jet), we would expect  corresponding differences in the $p_T$ of $b$-jet pair assigned to the $h$ boson. Likewise, we would expect the different $p_T$'s of the (prompt) $W^\pm$ bosons to transfer efficiently to the lepton spectra at the detector level, despite the dilution due to the fact the leptons emerging from the top (anti)quark leptonic decays, via (secondary) $W^\pm$'s, are rather identical. 

\begin{figure}[!htb]
	\begin{center}
		\includegraphics[width=0.45\linewidth]{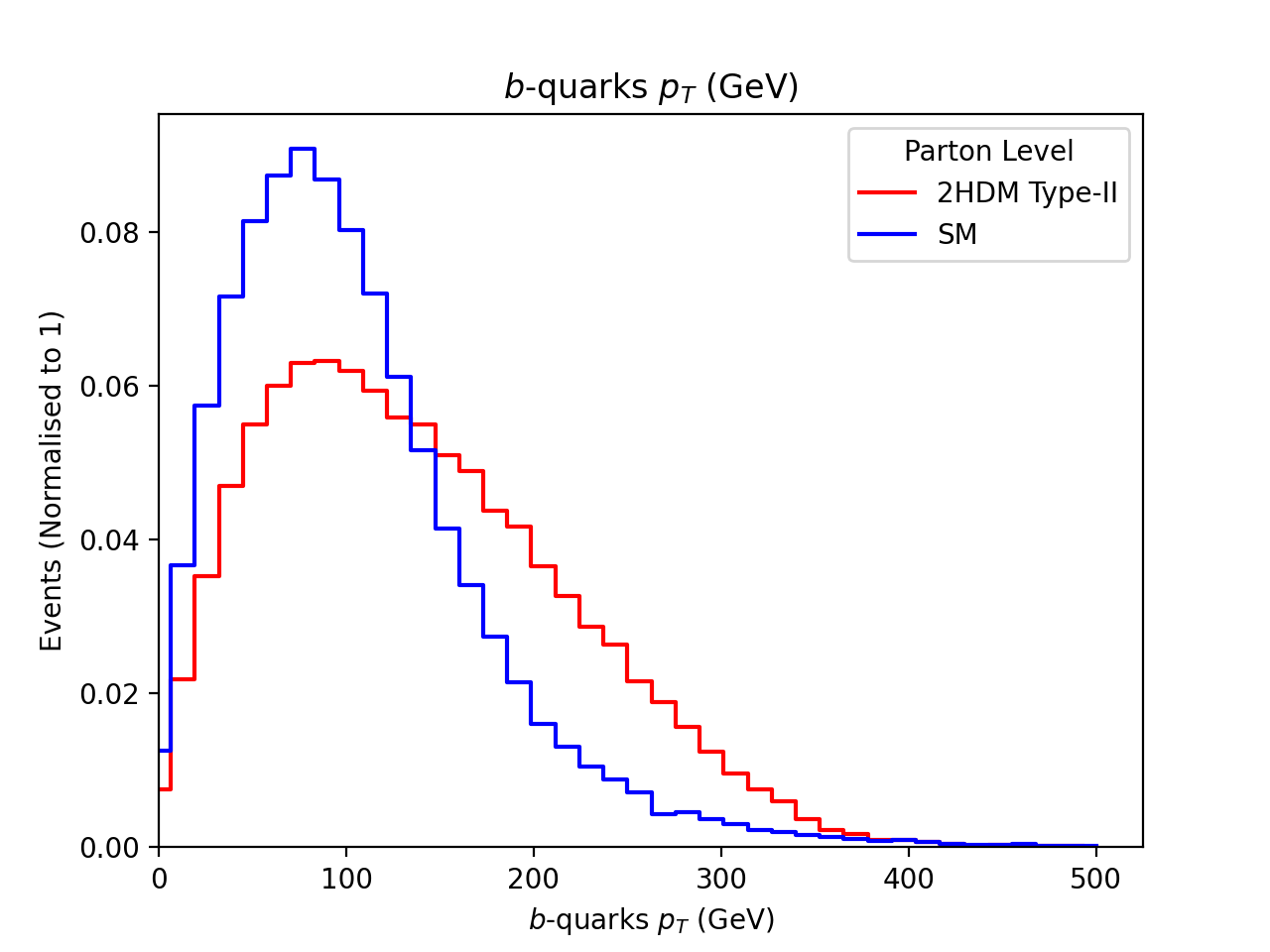}		
		\caption{\label{bquark_parton_level}
			Normalised transverse momentum distributions for all $b$-quarks at the parton level.  
		}
	\end{center}
\end{figure}

As a final parton level study, we also plot the $p_T$ distribution of the $b$-quarks for both models. This is done in Fig.~\ref{bquark_parton_level}. We can again see that there are differences  between the two theoretical scenarios: essentially, the  $b$-quarks have a wider range in  $p_T$’s in the 2HDM type-II when compared to the SM. Hence, we would expect the resulting $b$-jets to have a similar kinematic spread of $p_T$’s at the  detector level. 

\subsubsection{Hadron Level Analysis}
In this sub-section, we consider a detector level analysis of the hadronised events, for the purpose of identifying kinematics differences between the two theoretical scenarios considered, as a legacy of those just seen at parton level. 
\begin{figure}[H]
	\begin{center}
		\includegraphics[width=0.44\linewidth]{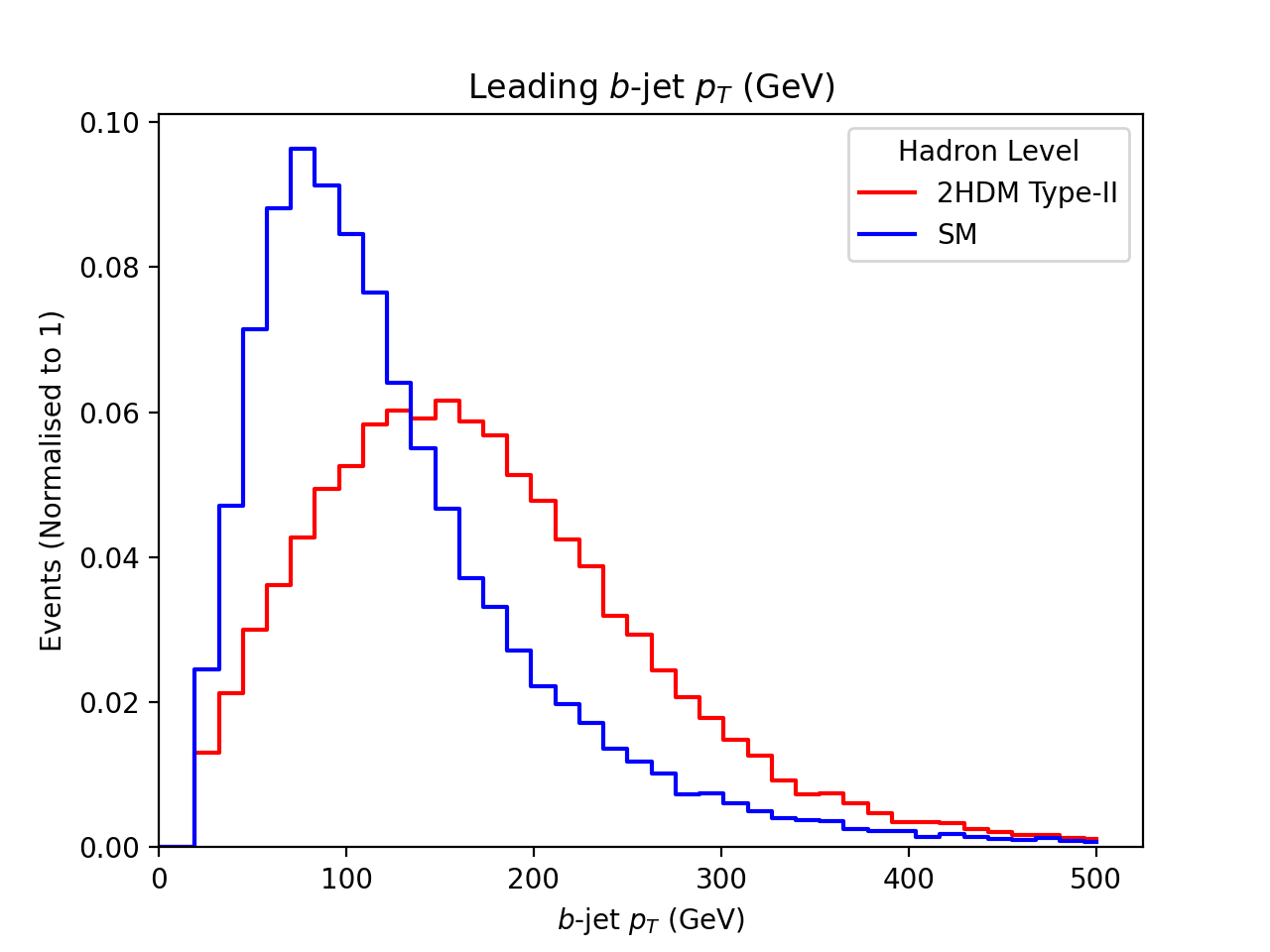}
		\includegraphics[width=0.44\linewidth]{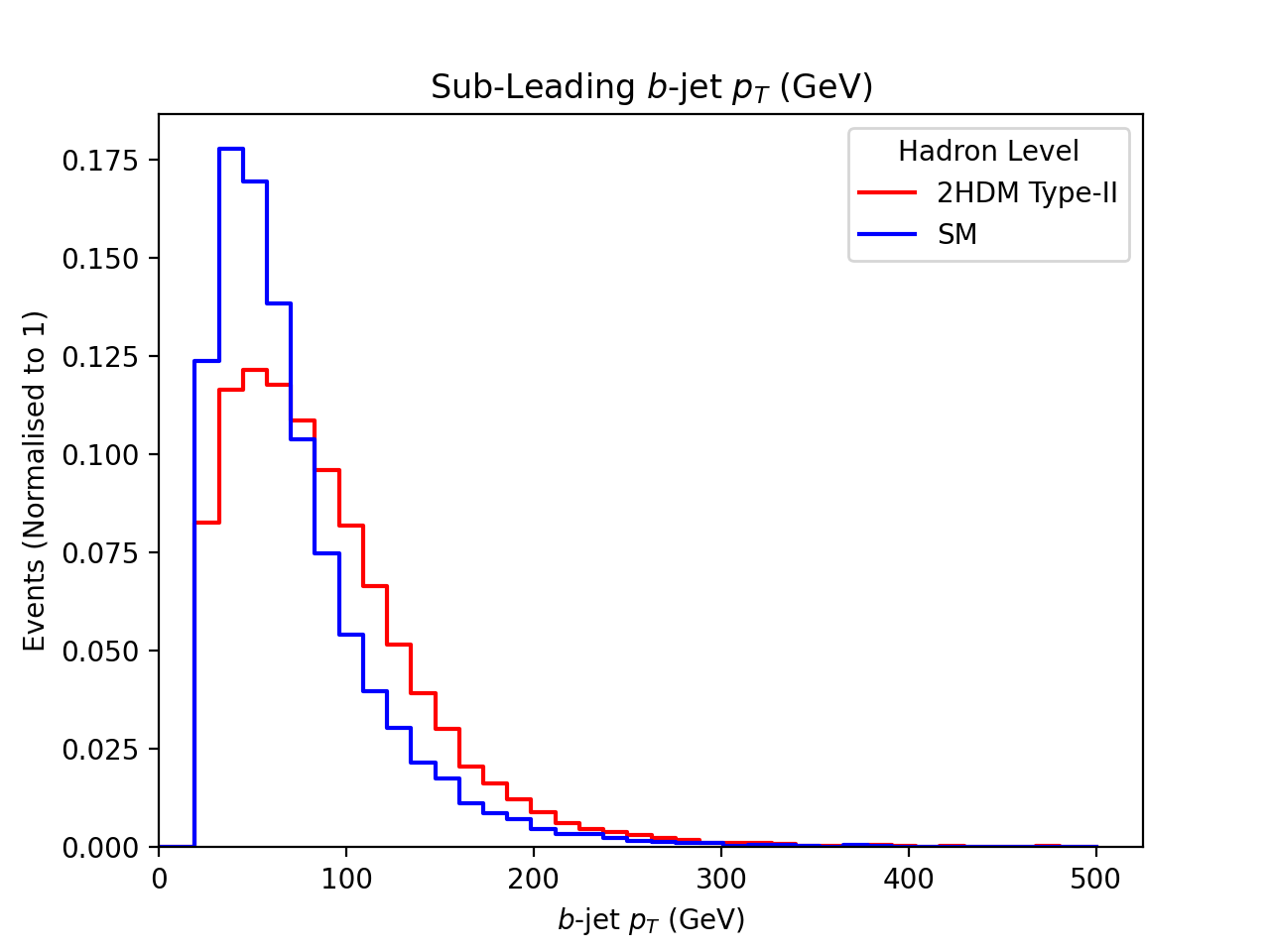}
		\\
		\includegraphics[width=0.44\linewidth]{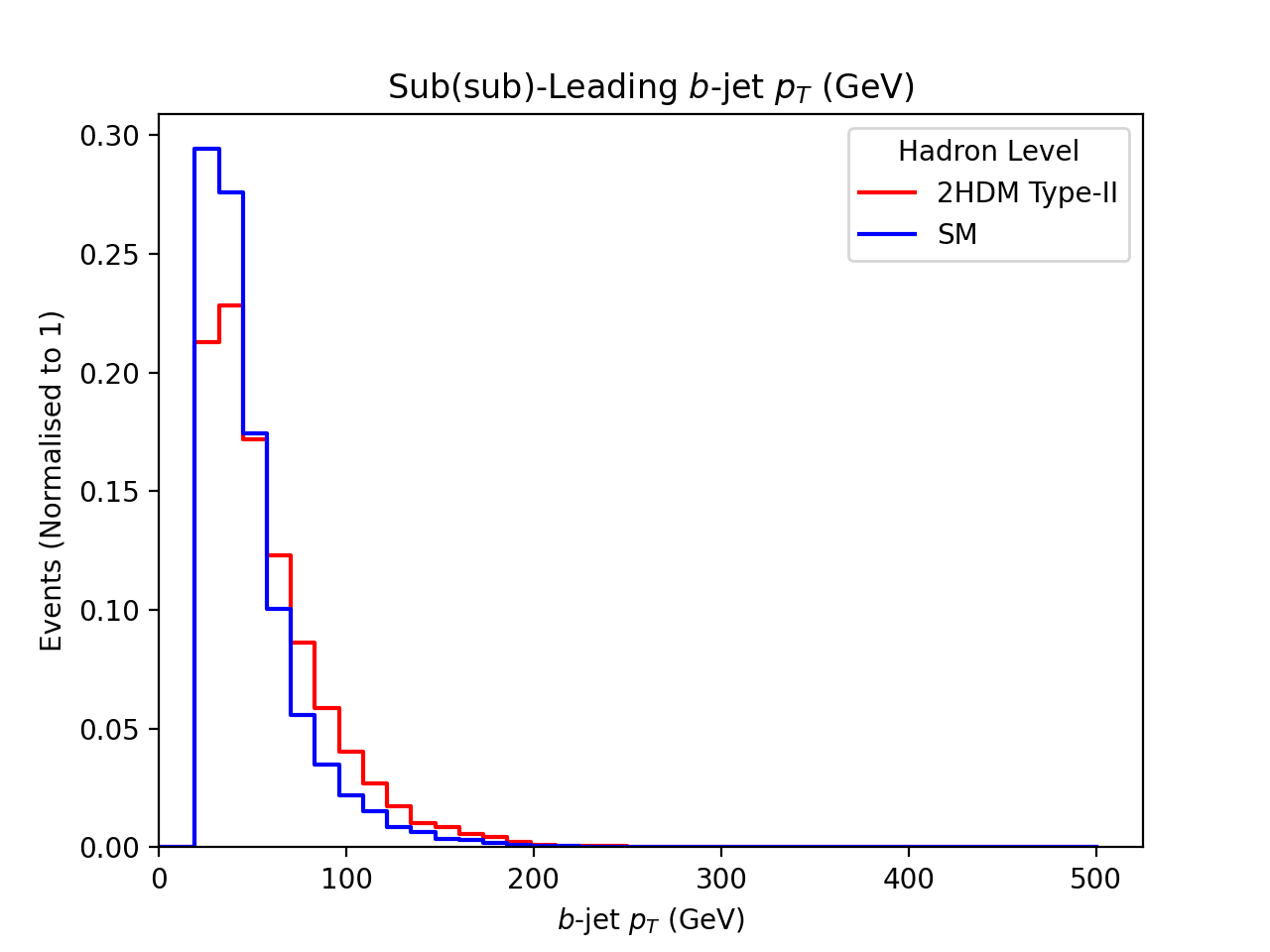}
		\caption{\label{hadron_level1}
			Upper panel: Normalised transverse momentum distributions of the leading $b$-jet (left) and sub-leading $b$-jet (right). Lower panel: Normalised ransverse momentum distributions of the sub-sub-leading $b$-jet.  
		}
	\end{center}
\end{figure}
To start, we plot the ordered $b$-jets transverse momenta of the three $b$-jets that we naturally expect to emerge from $h$ and $t$ (as well as $\bar t$) decays, see Fig.~\ref{hadron_level1}. It is clear that the leading $b$-jet $p_T$ coming from  the 2HDM type-II BP considered is quite different from and much harder than that produced in the SM. This can indeed be traced to the fact that the  $b$-(anti)quarks produced in $h$ decays have significantly higher momentum in the BSM scenario than in the SM. There is a small difference for sub-leading $b$-jet $p_T$ too while for the sub-sub-leading $b$-jet the $p_T$'s look more or less the same for both theoretical models.

\begin{figure}[H]
	\begin{center}
		\includegraphics[width=0.45\linewidth]{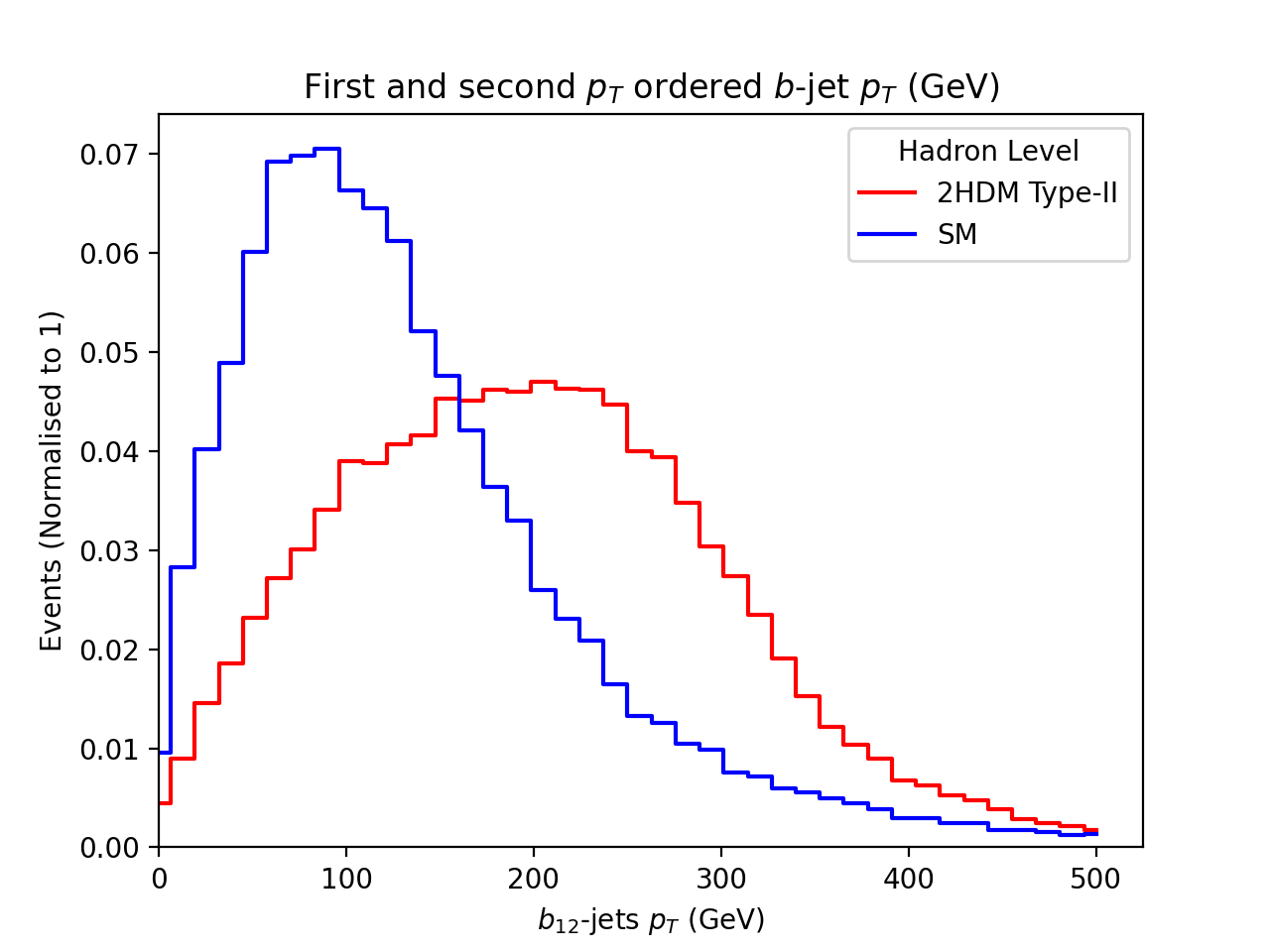}
		\includegraphics[width=0.45\linewidth]{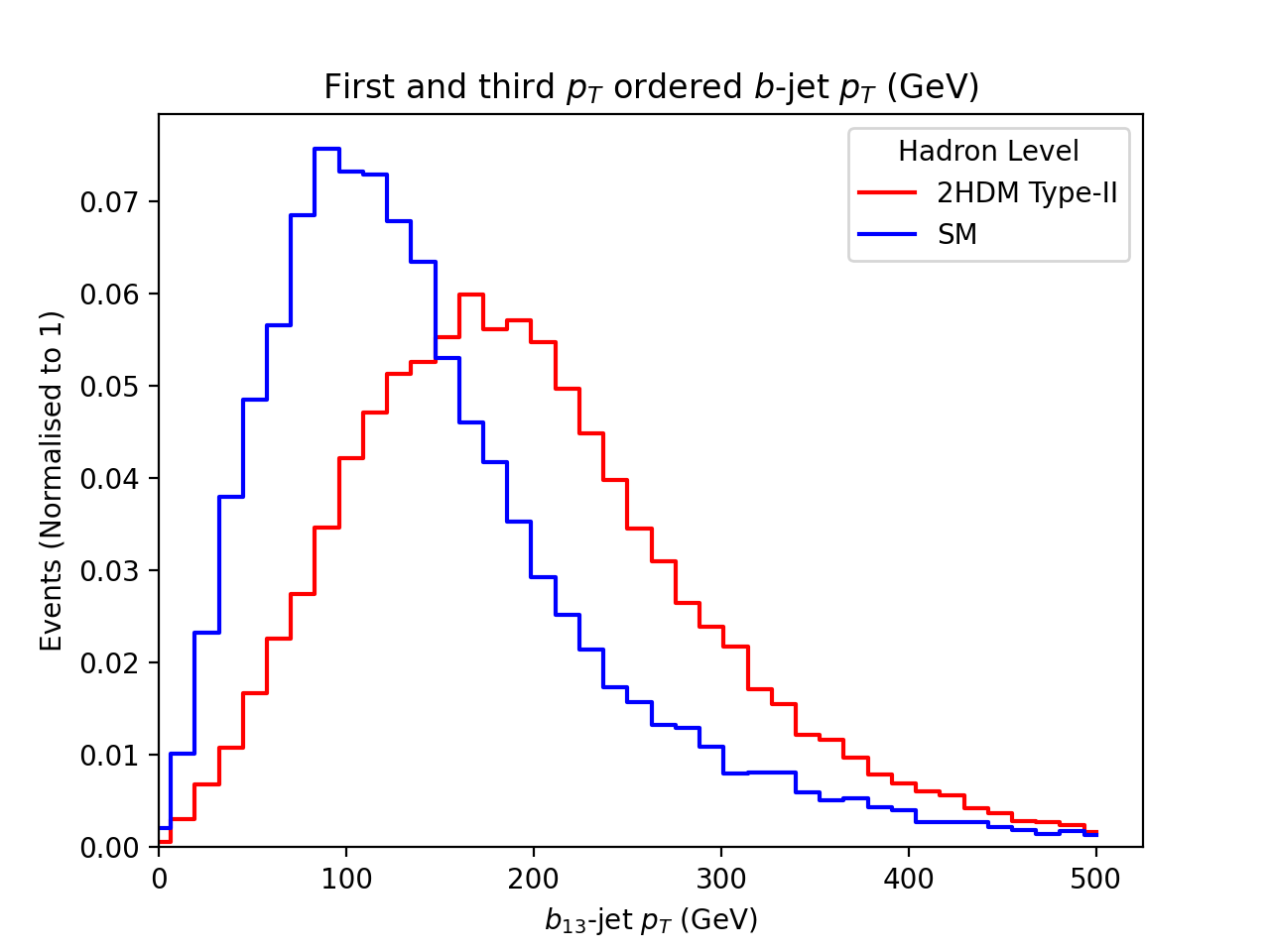}
		\\
		\includegraphics[width=0.45\linewidth]{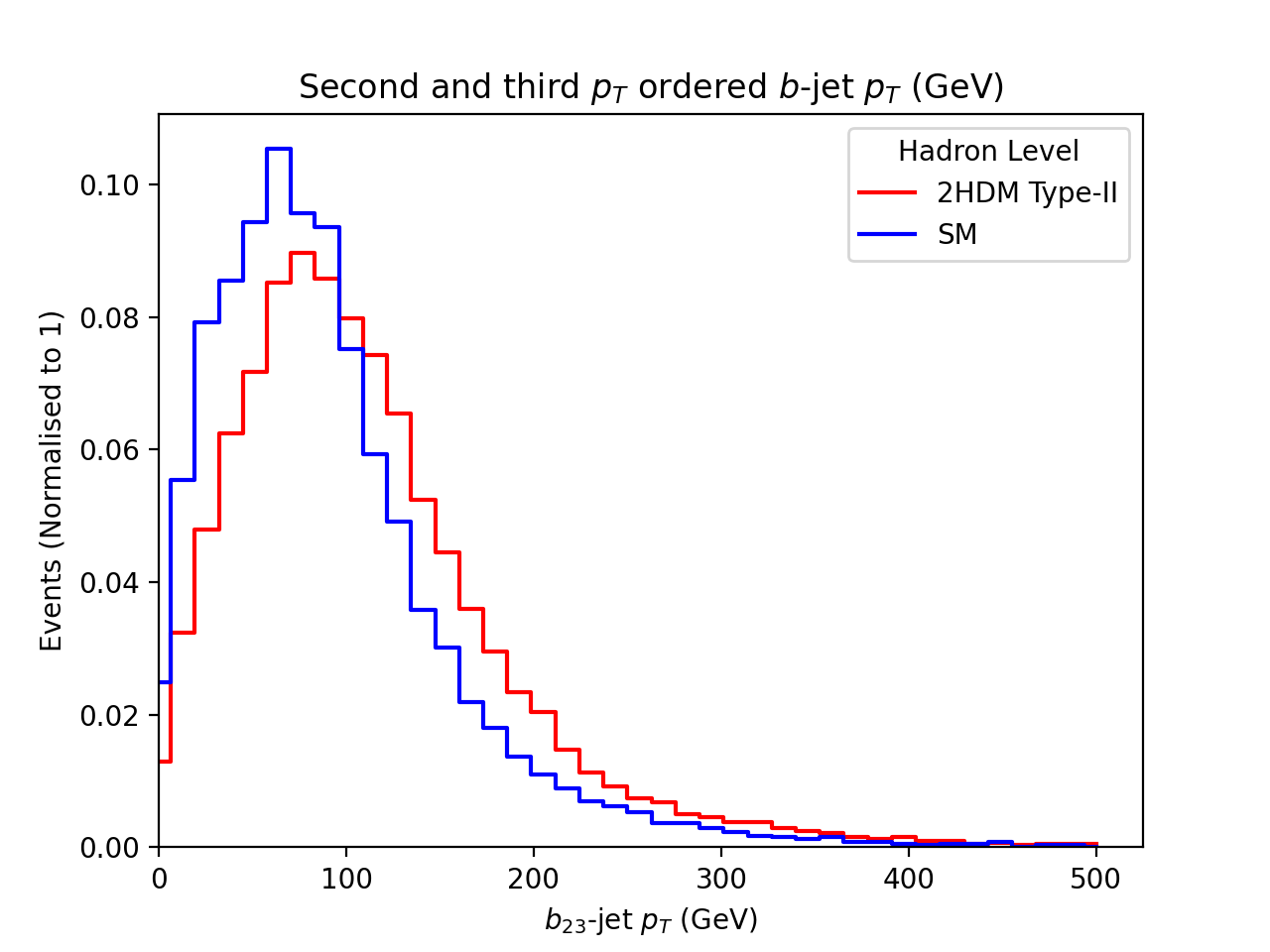}
		\caption{\label{hadron_level2}
			Upper panel: Normalised  transverse momentum distributions of the leading plus sub-leading$b$- jes pair, $b_{12}$ (left),  and  leading plus sub-sub-leading $b$-jet pair, $b_{13}$ (right). Lower panel: Normalised transverse momentum distribution of the sub-leading plus sub-sub-leading $b$-jet pair, $b_{23}$.}
	\end{center}
\end{figure}

As mentioned, our detector level events should have three $b$-jets with two of them coming from the $h$ decay and the third one coming from the top (anti)quark one. In Fig.~\ref{hadron_level2}, we plot the combined transverse momentum distributions of these three $b$-jets permutated in pairs. The $p_T$ for the  leading plus sub-leading $b$-jet pair  ($b_{12}$) and for the leading plus 
sub-sub-leading $b$-jet pair ($b_{13}$) are quite different for the two theoretical models. This is clearly due to the fact that, between the 2HDM type-II and SM, differences in the $p_T$ of the leading $b$-jet (as seen in Fig.~\ref{hadron_level1}) are more marked in comparison to those between the other two $b$-jets. For the same reason, we  observe only a slight difference in $p_T$ for the sub-leading plus sub-sub-leading $b$-jets pair ($b_{23}$).

\begin{figure}[H]
	\begin{center}
		\includegraphics[width=0.45\linewidth]{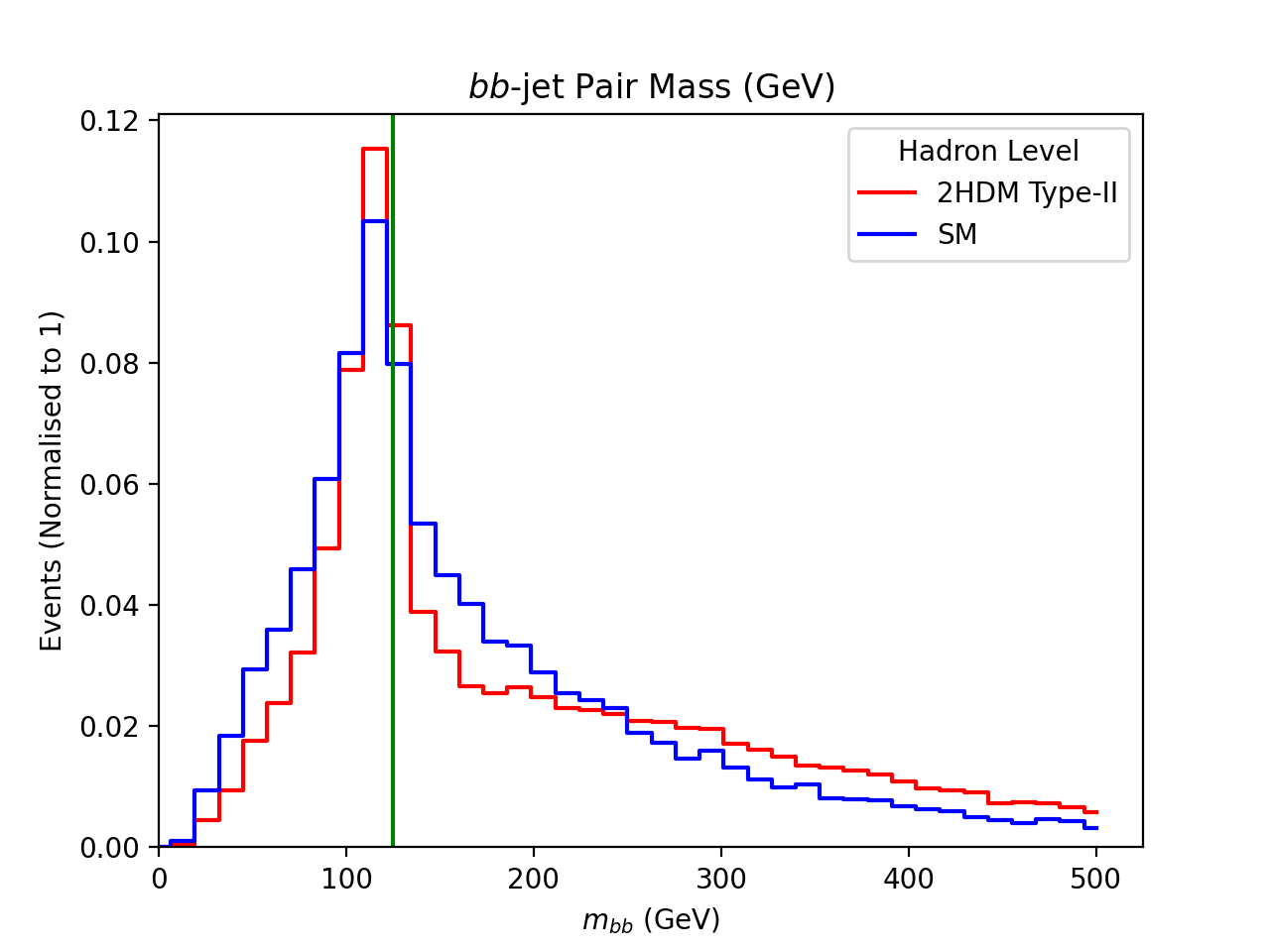}
		
		\caption{\label{hadron_level3} Normalised invariant mass distribution of all $b$-jet pairs.
	The vertical green line represents the MC truth value of the $h$ mass, $m_h=125$ GeV.	}
	\end{center}
\end{figure}

Next, we look at the invariant mass of all pairs of clustered $b$-jets, $m_{bb}$,  in order to reconstruct and identify the SM Higgs mass resonance. 
Remarkably, from Fig.~\ref{hadron_level3}, we can see that, in the 2HDM type-II framework, the mass reconstruction is somewhat sharper in the BSM scenario than it is in the SM, signaling that the effect of combinatorics is milder in the former case than in the latter. However, both distributions are somewhat dis-aligned from the MC truth value of the corresponding Higgs boon resonance. 

\begin{figure}[H]
	\begin{center}
		\includegraphics[width=0.45\linewidth]{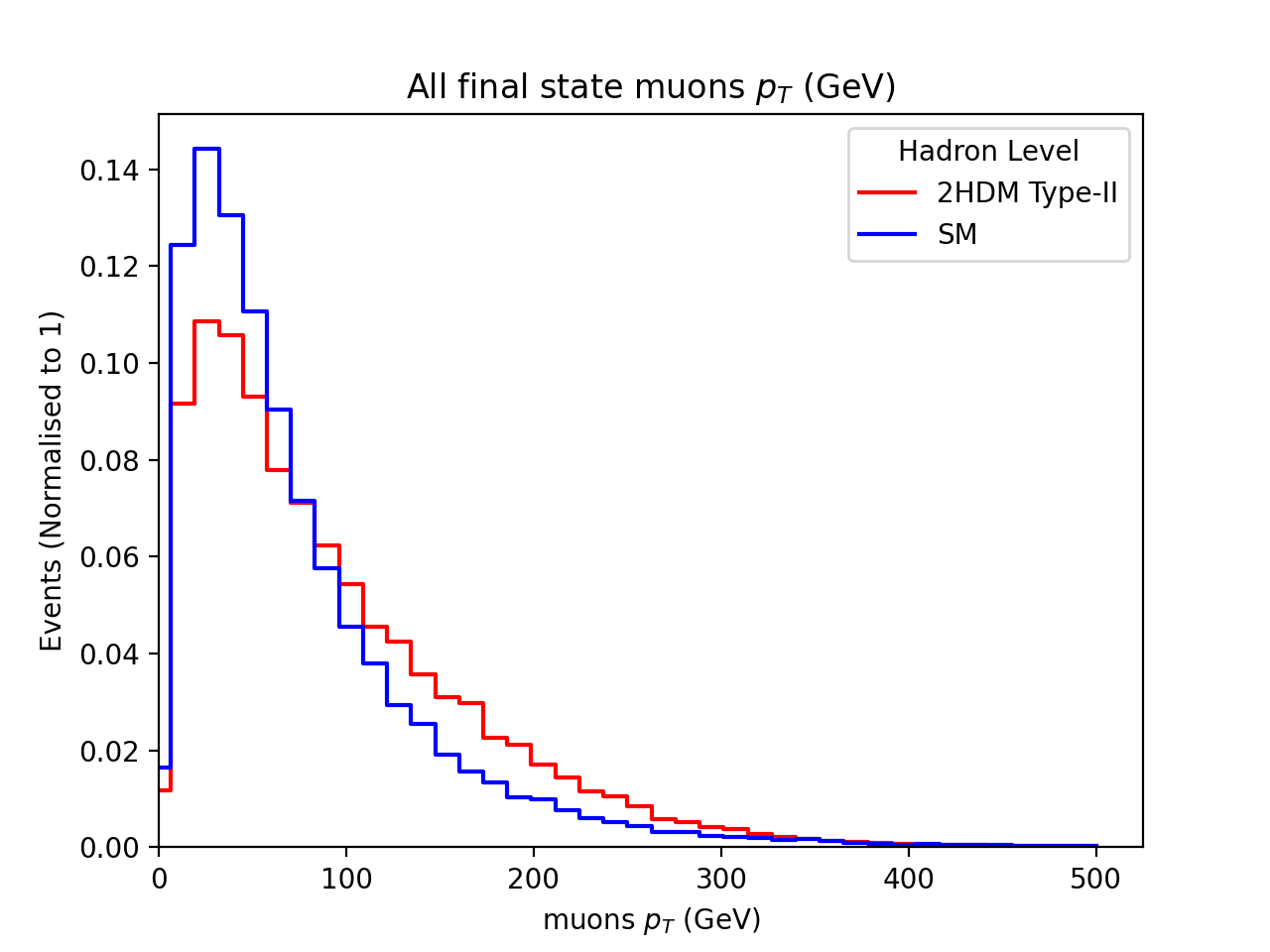}
		\includegraphics[width=0.45\linewidth]{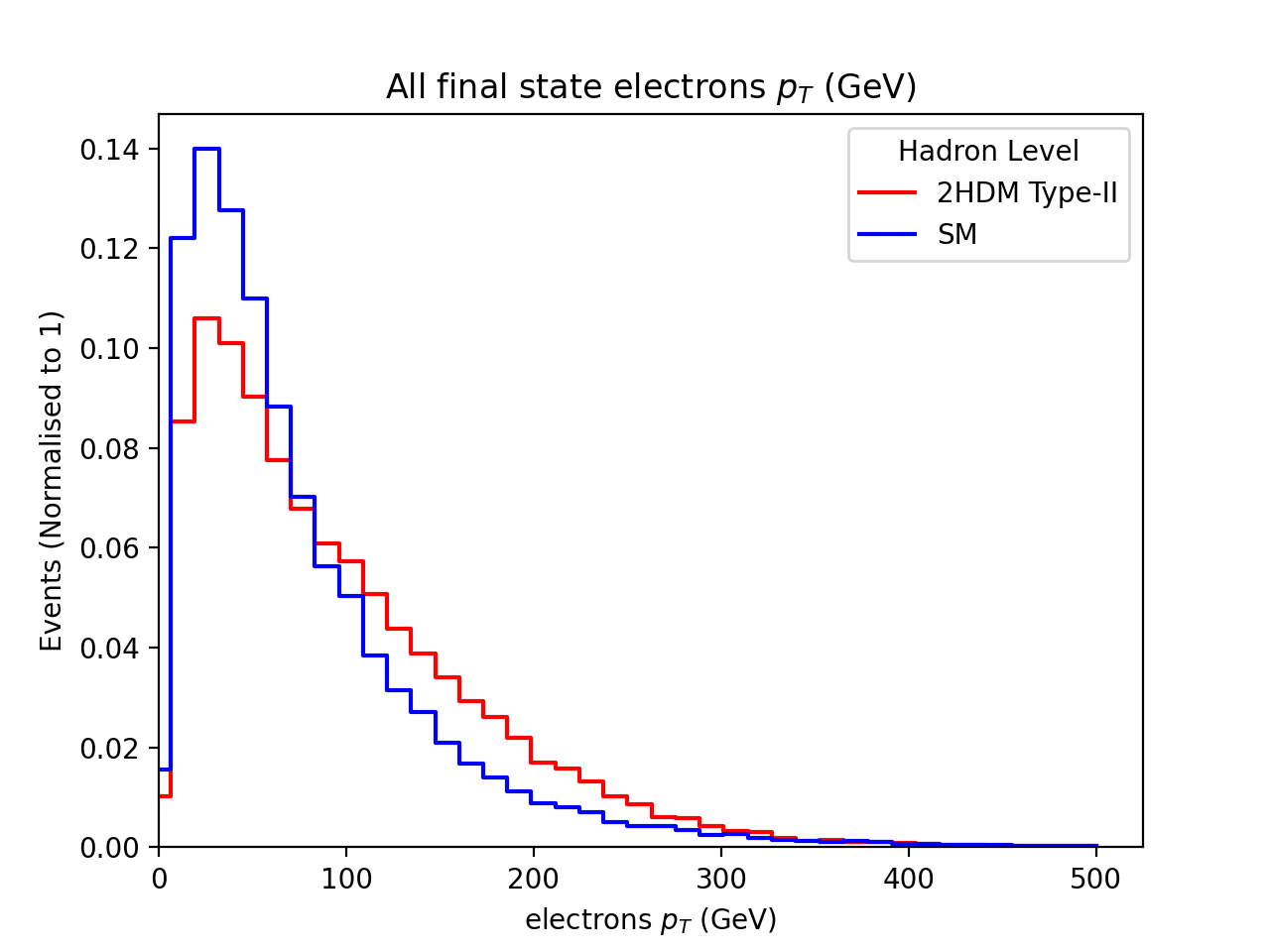}
		\caption{\label{hadron_level4}
			Normalised transverse momentum distributions of all muons (left) and electrons (right).}
	\end{center}
\end{figure}

Finally, we look at the transverse momentum distributions for all muons and electrons coming from top (anti)quark and $W^\pm$ boson decays, see  Fig.~\ref{hadron_level4}. There is a small change in the shape of these kinematic distributions between the two models, with a tendency for the 2HDM type-II events to be somewhat harder than the SM ones on this variable. 

Given that differences between the BSM scenario and SM in kinematical observables persist at hadron level, in our forthcoming analysis of signal versus background, we will design a suitable cutflow: on the one hand, aimed at preserving the regions of phase space where such differences are manifest while, on the other hand, suppressing as much as possible the background relatively to the signal.  
\subsubsection{Signal-to-background Analysis}
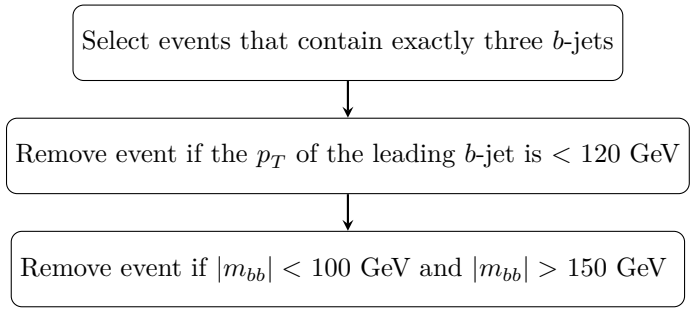
\begin{figure}[!htb]
	\centering
	
	\begin{tikzpicture}[node distance=1.5cm]
		\node (cut1) [node,align=center] {Select events that contain exactly three $b$-jets};
		\node (cut2) [node, align=center,below of=cut1] {Remove event if the $p_T$ of the leading $b$-jet is < 120  GeV};
		\node (cut3) [node, align=center,below of=cut2] {Remove event if $|m_{bb}|<$ 100  GeV and $|m_{bb}|>$ 150  GeV };
		\draw [arrow] (cut1) -- (cut2);
		\draw [arrow] (cut2) -- (cut3);

	\end{tikzpicture}
	\caption{Additional event selection used to compute the final significances of the signal.}
	\label{selection}
\end{figure}

As a final exercise, we attempt  to extract our signal based by enforcing the additional selection designed in  Fig.~\ref{selection} to finally compute the  signal-to-background significance rates for both  theoretical models considered. 

Thus, we calculate the expected event rates for the various signal and background processes, assuming an integrated luminosity of  ${\cal L}=$ $3000$ fb$^{-1}$ (corresponding to the one expected at the end of the  HL-LHC stage), as follows:
\begin{equation}
	N \ = \sigma \times \mathcal{L}.
\end{equation}
The event rates for the signal (in both models) and backgrounds are given in Tab.~\ref{tab:signalbackground2}.  It is clear that $gg, q\bar q  \to t\bar t$ is the dominant background followed by  $gg, q\bar q  \to t\bar t b \bar b $ and $gg, q\bar q  \to t\bar t h$. Contributions from the other backgrounds $gg,q\bar q \to t \bar t t \bar t$, $q\bar q \to W^+W^- h$, $q\bar q \to ZZh$ and  $q\bar q \to ZW^+W^-$ are instead negligible. 

The significance, $\Sigma$, is then calculated and is given by (as a function of signal ($S$) and background ($B$) event rates)
\begin{equation}
	\Sigma = \frac{N(S)}{\sqrt{N(B)}}.
\end{equation}

It is clear from Tab.~\ref{tab:signalbackground4} that the signal within the 2HDM type-II framework provides far better significances compared to the SM case. On the one hand, this means that the BSM signal will be seen much sooner at the HL-LHC than the SM one.
  On the other hand, given the differences identified between the two theoretical scenarios at the detector level in various observables, in turn stemming from rather different partonic behaviours, it may be possible to identify such a signal as being due to the 2HDM type-II. 
  
\begin{table}[H]
	\centering
	\hspace*{0.75truecm}
	\scalebox{1.2}{
		\begin{tabular}{|l|l|l|l|l|l|l|l|l|}
			\hline
			\multirow{1}{*}{Process} &
			\multicolumn{1}{c}{}\vline  \\   
			\hline
			bg (2HDM) &   279.624  \\
			\hline
			bg (SM) &   7.663  \\
			\hline
			$pp \rightarrow t\bar{t}$ & 9252.179   \\
			\hline
			$pp \rightarrow t\bar{t}h$ &110.127 \\
			\hline
			$pp \rightarrow t\bar{t}b\bar{b}$ &1013.760\\
			\hline
			$pp \rightarrow t\bar{t}t\bar{t}$ &0.182\\
			\hline
			$pp \rightarrow W^+W^-h$ &0.530\\
			\hline
			$pp \rightarrow ZZh$ &0.062\\
			\hline
			$pp \rightarrow ZW^+W^-$ &0.221\\
			\hline
			
		\end{tabular}
	}
	\caption{\label{tab:signalbackground2} Event rates of signal (in both models) and backgrounds for ${\cal L}=$ $3000$ fb$^{-1}$  upon enforcing all cuts.}
\end{table}

\begin{table}[H]
	\begin{center}
		\scalebox{1.2}{
			\begin{tabular}{ |c|c|c| }
				\hline
				& 2HDM  & SM     \\
				\hline
				${\cal L}=$ $3000$ fb$^{-1}$ &2.744  & 0.075   \\
				\hline
				
			\end{tabular}
		}
		\caption{\label{tab:signalbackground4} Final $\Sigma$ values calculated for ${\cal L}=$ $3000$ fb$^{-1}$  after enforcing all cuts.}
	\end{center}
\end{table}

This is clearly a preliminary conclusion, primarily aimed at alerting the experimental community to the fact that `SM-like Higgs boson production in association with single-top' through the bg channel can be used to test the presence of an extended Higgs sector. There is, in fact, much more that it could be done to put this on a more solid footing. For a start, it should  be noted that our analysis used a rather simple cut-and-count method to calculate the signal significance and only looked at one final state of the signal  (the decay $h\to b\bar b$). Using a more advanced method for extracting the signal significance, like a maximum-likelihood fit or machine learning approaches,  as well as analysing more signal final states, by considering alternative $h$ and $W$ decays, we ultimately  anticipate a substantial boost to the expected signal significance.
\section{Summary}
\noindent
In conclusion, 
we have investigated the parameter spaces of the 2HDM type-I as well as II and found that, while the type-I does not appear to contain any significantly larger cross-sections than in the SM for `SM-like Higgs boson production in association with single-top',  
the type-II contains many. Herein, parameter space points were found for both the so-called `wrong-sign solution' (of the bottom (anti)quark Yukawa coupling) and `alignment limit'   the vast majority of which have a cross-section far greater than that the one expected in the SM, albeit limitedly to the bg sub-process, most notably so in the former than the latter parameter space configuration. A representative BP was then selected over 
the region of 2HDM type-II parameter space realising the `wrong-sign solution' 
for detailed MC analysis in presence of the most significant backgrounds (carried out in parallel to the SM case). This implemented a cut-and-count approach at the detector level, which has lead two a twofold result. First, the 2HDM type-II signal may be established at the HL-LHC much before the SM one. Second, the corresponding excess events in the 2HDM type-II  would have kinematic features notably different from the SM case, thus offering a diagnostic scope of the underlying Higgs dynamics.

Our analysis was a preliminary one aimed at alerting the experimental community to the potential of the bg sub-process 
triggering `SM-like Higgs boson production in association with single-top' to test a possible non-standard nature of EWSB, better than the alternative two channels (bq and qq) can do, as their production cross-sections are essentially the same in both theoretical scenarios considered. In fact, so far, only inclusive approaches have been adopted in pursuing this signature, i.e., capturing all such three sub-processes simultaneously, an approach that may better be dismissed in the future, at least in the search for a very peculiar  configuration of the 2HDM type-II, the aforementioned `wrong-sign solution',  which has survived precise experimental scrutiny of the SM-like Higgs boson signals to date.

\section*{Acknowledgements}
\noindent
SM is supported in part through the NExT Institute and the STFC Consolidated Grant ST/L000296/1. SJ is partially funded by DISCnet studentship. We thank Souad Semlali for the provision of the 2HDM type-I model file used in this paper. CB and SJ acknowledge the use of the IRIDIS5 High-Performance Computing Facility, and associated support services at the University of Southampton, in the completion of this work.

\bibliographystyle{ieeetr}
\bibliography{BibliographyFile}

\end{document}